\documentclass{aa}
\usepackage{graphicx}
\usepackage[]{natbib}
\begin{document}
   \title{Monitoring the hard X-ray sky with SuperAGILE}

   \author{
    M. Feroci$^{1}$, E. Costa$^{1}$, E. Del Monte$^{1}$, I. Donnarumma$^{1}$,
    Y. Evangelista$^{1}$, I. Lapshov$^{1,15}$, F. Lazzarotto$^{1}$,
    L. Pacciani$^{1}$, M. Rapisarda$^{12}$, P. Soffitta$^{1}$,
    G. Di Persio$^{1}$, M. Frutti$^{1}$, M. Mastropietro$^{24}$,
   E. Morelli$^{8}$,  G. Porrovecchio$^{1}$,    A. Rubini$^{1}$,
A. Antonelli$^{18}$, A. Argan$^{1}$,  G. Barbiellini$^{4,5,3}$, F.
Boffelli$^{13}$, A. Bulgarelli$^{8}$, P. Caraveo$^{6}$, P. W.
Cattaneo$^{13}$, A. W. Chen$^{6}$, V. Cocco$^{2}$, S.
Colafrancesco$^{14,17}$,  S. Cutini$^{14}$, F. D'Ammando$^{1,2}$,
 G. De Paris$^{1}$, G. Di Cocco$^{8}$,   G.
Fanari$^{14}$,  A. Ferrari$^{3,16}$, M. Fiorini$^{6}$, F.
Fornari$^{6}$, F. Fuschino$^{8}$, T. Froysland$^{3,7}$,  M.
Galli$^{9}$, D. Gasparrini$^{14}$, F. Gianotti$^{8}$, P.
Giommi$^{14,17}$, A. Giuliani$^{6}$, C. Labanti$^{8}$,  F.
Liello$^{5}$, P. Lipari$^{10,11}$, F. Longo$^{4,5}$, E.
Mattaini$^{6}$, M. Marisaldi$^{8}$, A. Mauri$^{8}$, F.
Mauri$^{13,\dag}$, S. Mereghetti$^{6}$,  E. Moretti$^{4,5}$, A.
Morselli$^{7}$, A. Pellizzoni$^{19}$, F. Perotti$^{6}$, G.
Piano$^{1,2,7}$, P. Picozza$^{2,7}$, M. Pilia$^{19}$, C.
Pittori$^{14}$, C. Pontoni$^{3,5}$,  B. Preger$^{14}$, M.
Prest$^{5}$, R. Primavera$^{14}$, G. Pucella$^{12}$,  A.
Rappoldi$^{13}$, E. Rossi$^{8}$,  S. Sabatini$^{2}$, P.
Santolamazza$^{14}$,  M. Tavani$^{1,2,7,3}$, S. Stellato$^{14}$,
F. Tamburelli$^{14}$, A. Traci$^{8}$, M. Trifoglio$^{8}$, A.
Trois$^{1}$, E. Vallazza$^{5}$, S. Vercellone$^{20}$, F.
Verrecchia$^{14}$, V. Vittorini$^{1,3}$, A. Zambra$^{3,6}$, D.
Zanello$^{10,11}$, and L. Salotti$^{17}$  }

\institute{
$^1$ INAF-IASF Roma, via del Fosso del Cavaliere 100, I-00133 Roma, Italy\\
$^2$ Dipartimento di Fisica, Universit\'a Tor Vergata, via della Ricerca Scientifica 1,I-00133 Roma, Italy\\
$^3$ Consorzio Interuniversitario Fisica Spaziale (CIFS), villa Gualino - v.le Settimio Severo 63, I-10133 Torino, Italy\\
$^4$ Dip. Fisica, Universit\`a di Trieste, via A. Valerio 2, I-34127 Trieste, Italy\\
$^5$ INFN Trieste, Padriciano 99, I-34012 Trieste, Italy\\
$^6$ INAF-IASF Milano, via E. Bassini 15, I-20133 Milano, Italy\\
$^7$ INFN Roma Tor Vergata, via della Ricerca Scientifica 1, I-00133 Roma, Italy\\
$^8$ INAF-IASF Bologna, via Gobetti 101, I-40129 Bologna, Italy\\
$^9$ ENEA Bologna, via don Fiammelli 2, I-40128 Bologna, Italy\\
$^{10}$ INFN Roma 1, p.le Aldo Moro 2, I-00185 Roma, Italy\\
$^{11}$ Dip. Fisica, Universit\`a La Sapienza,p.le Aldo Moro 2, I-00185 Roma, Italy\\
$^{12}$ ENEA Frascati, via Enrico Fermi 45, I-00044 Frascati(RM), Italy\\
$^{13}$ INFN Pavia, via Bassi 6, I-27100 Pavia, Italy\\
$^{14}$  ASI Science Data Center, ESRIN, I-00044 Frascati(RM), Italy\\
$^{15}$ IKI, Moscow, Russia\\
$^{16}$ Dipartimento di Fisica, Universita\' di Torino, Torino, Italy\\
$^{17}$  Agenzia Spaziale Italiana, viale Liegi 26 , 00198 Roma, Italy\\
$^{18}$ Osservatorio Astronomico di Roma, Via di Frascati 33, I-00040 Monte Porzio Catone, Italy \\
$^{19}$ Osservatorio Astronomico di Cagliari, loc. Poggio dei Pini, strada 54, I-09012, Capoterra (CA) \\
$^{20}$ INAF-IASF Palermo, Via Ugo La Malfa 153, 90146 Palermo,
Italy\\
$^{\dag}$ Deceased }


   \offprints{M. Feroci \email{marco.feroci@iasf-roma.inaf.it}}

%

   \date{}


\abstract
{SuperAGILE is the hard X-ray monitor of the AGILE gamma ray
mission, in orbit since 23$^{rd}$ April 2007. It is an imaging
experiment based on a set of four independent silicon strip
detectors, equipped with one-dimensional coded masks, operating in
the nominal energy range 18-60 keV.}
{The main goal of SuperAGILE is the  observation of cosmic sources
simultaneously with the main gamma-ray AGILE experiment, the Gamma
Ray Imaging Detector (GRID). Given its $\sim$steradian-wide field
of view and its $\sim$15 mCrab day-sensitivity, SuperAGILE is also
well suited for the long-term monitoring of Galactic compact
objects and the detection of bright transients.}
{The SuperAGILE detector properties and design allow for a 6
arcmin angular resolution in each of the two independent
orthogonal projections of the celestial coordinates. Photon by
photon data are continuously available by the experiment
telemetry, and are used to derive images and fluxes of individual
sources, with integration times depending on the source intensity
and position in the field of view.}
{In this paper we report on the main scientific results achieved
by SuperAGILE over its first two years in orbit, until April 2009.
The scientific observations started on mid-July 2007, with the
Science Verification Phase, continuing during the complete AGILE
Cycle 1 and the first $\sim$half of Cycle 2. Despite of the
largely non-uniform sky coverage, due to the pointing strategy of
the AGILE mission, a few tens of Galactic sources were monitored,
sometimes for unprecedently long continuous periods, detecting
also several bursts and outbursts. Approximately one gamma ray
burst per month was detected and localized, allowing for prompt
multi-wavelength observations. Few extragalactic sources in bright
states were occasionally detected as well. The light curves of
sources measured by SuperAGILE are made publicly available on the
web in nearly realtime. For a proper scientific use of these we
provide the reader with the relevant scientific and technical
background. }
{}


\authorrunning {M. Feroci et al.}
\titlerunning {Hard X-ray sky with SuperAGILE}
\maketitle
%

\section{Introduction}

The SuperAGILE experiment (Feroci et al. 2007) is the hard X-ray
imager of the Italian gamma-ray astronomy mission AGILE (Tavani et
al. 2009). The primary instrument of the AGILE payload is the GRID
(Gamma Ray Imaging Detector, Barbiellini et al. 2002, composed of
a Silicon Tracker (ST, Prest et al. 2003) and a small CsI(Tl)
"mini"-calorimeter (MCAL, Labanti et al. 2009), surrounded by a
plastic anticoincidence system (ACS, Perotti et al. 2006).

The main scientific goal of AGILE is the observation of the
gamma-ray sky in the energy range from 50 MeV to a few GeV, over a
field of view in excess of 2 steradian, ten years after the demise
of the EGRET experiment onboard the Compton Gamma Ray Observatory.
Several classes of celestial sources are known to emit gamma-rays
in this energy band, most notably rotation powered pulsars, active
galactic nuclei (especially blazars), supernova remnants and,
occasionally, gamma-ray bursts. Other types of sources, such as
galactic microquasars, low and high mass X-ray binaries or
magnetars were expected to possibly emit gamma-rays in some
special conditions (e.g., from shocks in the jet or in the
accretion flow), but none of them had been unambiguously
classified as a gamma ray source prior to the recent observations
by AGILE or Fermi-GLAST (with the exception of the debated cases
of LS 5039 and LSI +61 303).

The SuperAGILE experiment was designed to guarantee the hard X-ray
(18-60 keV) monitoring in the tens of mCrab sensitivity range of
the central $\sim$steradian of the GRID field of view, aiming at
identifying in the hard X-rays sources and/or conditions of
special interest to the GRID, providing accurate ($\sim$arcmin)
localizations and allowing for multi-band observations. In
addition to that, SuperAGILE independently acts as a wide field
monitor for bright galactic sources, with photon by photon
transmission and 5-$\mu$s time accuracy.

A detailed description of the SuperAGILE experiment (pre-flight)
may be found in Feroci et al. (2007), while information about the
early in-flight operation and technical performance may be found
in Feroci et al. (2008). Here we summarize the main instrument
properties relevant to the aim of the present paper, that is
providing an overview of the main scientific results achieved by
the experiment during the first 20 months of observations from
space, as well as guidelines and limits for a proper scientific
use of the SuperAGILE data.

\begin{figure}
\centering
\includegraphics[width=9cm, clip]{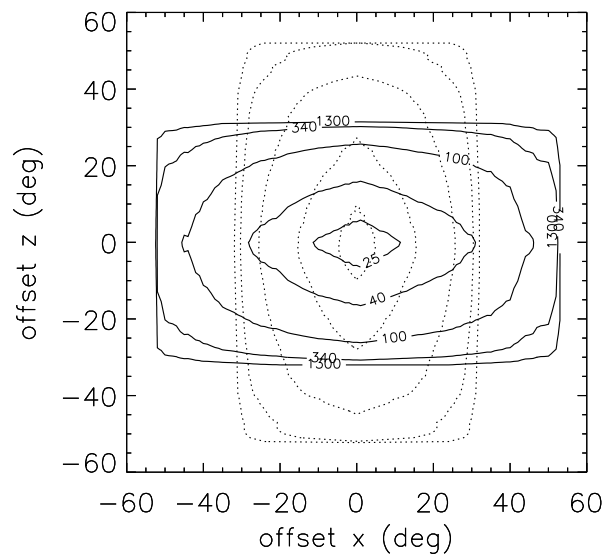}
 \caption{Representation of the field of view of the two pairs of
 SuperAGILE detectors, encoding two directions in the sky.
 Here we show with solid lines the sensitivity contours of
one pair of detectors (i.e., one coding direction, either X or Z).
The contours in dotted lines represent the sensitivity of the
other pairs of detectors, rotated by 90$^{\circ}$, providing the
other coordinate. Sensitivity is given in mCrab (20-60 keV) at
5-$\sigma$ confidence level, over a 50 ks integration time, for
one direction only (half of the experiment area). }
\label{fig_fov}
\end{figure}

SuperAGILE is a coded-mask experiment devoted to image the
transient hard X-ray sky by means of a set of four one-dimensional
Silicon microstrip detectors, encoding the sky in two orthogonal
directions. The field of view (FOV) of each pair of detectors is
rectangular, 107$^{\circ}$ x 68$^{\circ}$ at zero response (Fig.
\ref{fig_fov}). Thus, the central 68$^{\circ}$ x 68$^{\circ}$
region is encoded in both directions (2 x 1D), while the
peripheral $\sim\pm$20$^{\circ}$ of the FOV are coded in one
dimension only. Due to severe limitations at design and operation
level (see Feroci et al. 2007, 2008a), SuperAGILE operates in the
non-optimal (for a Silicon detector) energy range from 18 to 60
keV, being noise-limited at the low energy bound and efficiency
limited at the upper bound, and behind a 5 mm thick shield of
plastic scintillator (part of the AGILE ACS). The energy
resolution is $\sim$8 keV Full Width at Half Maximum,
energy-independent. The one-dimensional point spread function
(PSF) is 6 arcmin on-axis and the point source location accuracy
reaches 1-2 arcmin for detections above $\sim$10$\sigma$,
including a $\sim$1 arcmin systematics contribution from the AGILE
satellite attitude reconstruction. The SuperAGILE on axis
5-$\sigma$ sensitivity is $\sim$20 mCrab for a 50 ks net exposure
(independently in each direction, that is half of the area of the
experiment). Due to the triangular response of the collimator, the
FOV at half sensitivity with 2 x 1D imaging is approximately
30$^{\circ}$ x 30$^{\circ}$. Instead, in each direction the
sensitivity is better than 100 mCrab (5$\sigma$, 50 ks, two
detectors only) within the central 40$^{\circ}$ x 80$^{\circ}$
region (see Fig.\ref{fig_fov}). The on-axis effective area is 250
cm$^{2}$ (total of 4 detectors, at 20 keV). The experiment
background is dominated by the diffuse X-rays. Particle induced
background is largely discriminated in amplitude and by the
anticoincidence logic with the ACS, that actually causes a large
dead-time effect when crossing the south atlantic anomaly,
preventing in practice scientific observations in this orbital
phase (see, e.g., Feroci et al. 2008a).

The SuperAGILE experiment entered its nominal observing phase on
mid-July 2007, after the successful completion of its
Commissioning Phase. Due to the inaccessibility of the Crab Nebula
in June, caused by the  Sun constraints of the spacecraft, the
first light of the experiment was taken on the Vela region, with
the detection of the galactic binary system GX 301-2. Between July
and October 2007 the experiment was calibrated in flight, using
the Crab Nebula to scan several positions in the FOV, as a part of
its Science Verification Phase. The latter was completed by the
end of November 2007 with the observation of the Vela field, the
Galactic Center and the Cygnus region. On December 2007 the AGILE
Observing Cycle 1 started, with typical continuous exposures of 4
weeks to the same field. Contrary to those of the GRID, the
SuperAGILE data of Cycle 1 are not open to Guest Observers.
However, light curves of the sources detected by SuperAGILE are in
public distribution in nearly realtime through a web page (see $\S
\ref{sa_web}$).

In the following sections we will describe the scientific data
products of SuperAGILE ($\S$ \ref{software}), review the status of
experiment calibration and pointing strategy ($\S$ \ref{calib} and
$\S$ \ref{pointings}), and overview the main scientific
observations carried out over the first $\sim$20 months of nominal
operations in orbit ($\S$ \ref{science}), including a preliminary
list of the detected sources.

\section{The SuperAGILE data analysis and scientific products}
\label{software}

The SuperAGILE experiment (hereafter SA)  transmits
photon-by-photon data to the AGILE telemetry and then to ground.
For each detected event information is available about its time of
arrival (with 2$\mu$s resolution and 5$\mu$s absolute accuracy),
energy (64 channels, formally between 0 and $\sim$90 keV) and
position on the detectors (strip address between 1 and 6144,
including the detector information).

The SA telemetry is downloaded from the spacecraft to the ground
station in Malindi (Kenya) once per orbit (approximately 100
minutes long). It is then transmitted to the Telespazio Fucino
Space Center and to the ASI Science Data Center in Frascati, where
it is both locally processed and transmitted to the SuperAGILE
Team in Rome (see Trifoglio et al. 2008 for a detailed description
of the AGILE data flow and processing). An automatic data
processing pipeline, SASOA (Lazzarotto et al. 2008, 2009),
extracts high-level products and scientific information from the
SA data. One dimensional  images of the sky in each of the two SA
coordinates are created using the data of the entire orbit. This
is aimed to derive an unbiased "picture" of the hard X-ray sky in
the experiment field of view, during that orbit. The results of
this analysis are automatically stored in a database, from where
orbital light curves of the sources positively detected can be
retrieved over the entire mission lifetime. The same procedure
runs automatically on a daily timescale, with the same approach.

Refined analysis can be performed by the user on sources of
interest. It consists in running the same procedures mentioned
above, but applying specific time and/or coordinates and/or energy
selections. For individual sources optimized results are obtained,
for example, by selecting the SA data on the time periods when the
source of interest was properly exposed. To these, further time
and energy selections can be applied, resulting in energy-selected
light curves of the source. Energy spectra can also be obtained by
running the imaging procedure iteratively on pre-determined energy
bins, and extracting for each of them the source intensity from
the combined images in the X and Z detectors. The results are then
stored in a standard fits \textit{pha} file, to be used in the
XSPEC spectral analysis package, together with the proper
instrument response matrix. The latter is accumulated for each
pair of detectors (X and Z). The response matrix is dependent on
the source position in the FOV and it is then computed for every
specific spectrum.

A crucial step in the SA
scientific analysis is the attitude correction. In fact, the
one-dimensional SuperAGILE images are severely affected by the
satellite attitude variations, as large as 0.1$^{\circ}$/s,
meaning that the satellite pointing direction varies by an amount
equal to the SA angular resolution every second. This is not
similarly critical for the GRID, that has full 2D imaging
capability and an angular resolution in the range of
$\sim$4$^{\circ}$. The detected position of every SA event needs
to be corrected to its expected location using the attitude
information from the AGILE star sensors, available in telemetry
with a 10 Hz frequency and a 1 arcmin accuracy, on each of the
three satellite axes.

\begin{figure}
\centering
\includegraphics[width=9.4cm,clip]{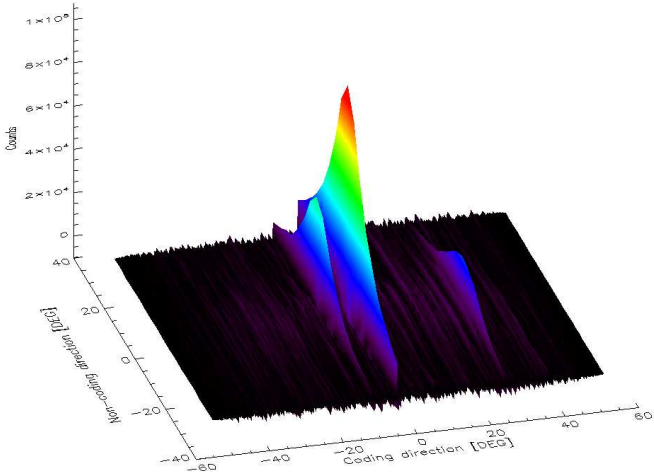}
\caption{A bi-dimensional image obtained by the SuperAGILE Z
detectors observing the Vela field on DOY 14, 2009. The Coding
direction shows the typical arcmin-resolution, while the
Non-Coding direction displays the counts obtained under different
attitude corrections (see text for details). The peak positions in
the Non-coding direction identify the source position along that
coordinate. The three peaks correspond to the sources 1A 1118-61
(brightest), GX 301-2 (second brightest, on the left), and Vela
X-1 (dimmest, on the right).} \label{fig_multi_emi}
\end{figure}

In a coded mask experiment, the attitude correction of an
individual photon must make an assumption on its incoming
direction. Unlike one would intuitively expect for a
one-dimensional imager like SA, the one-dimensional shift
compensating an attitude variation depends not only on the
projection of the source position along the coding direction, but
also along its non-coding direction. In practice, the attitude
variation provides an additional encoding to the photons, also in
the direction where the detector has no imaging capabilities. This
implies that the \textit{a-priori} assumptions on the incoming
direction of the photon (and the attitude correction thereof) and
the search for point sources in the FOV must be done in a
two-dimensional space, instead of the more "natural"
one-dimension. One of the two directions ("coding") has a
resolution of 3 arcminutes, as determined by the size of the sky
pixel of the experiment. The second one ("non-coding") has a
resolution (hereafter "zone") determined by a trade-off between
the sensitivity of the SA point spread function to the attitude
variation and the processing time. In the current version of the
SA scientific software, the angular size of a non-coding zone is
set to 4$^{\circ}$ ($\pm$2$^{\circ}$). A positive consequence is
that each SA detector becomes a two-dimensional imager, but at the
expenses of the signal-to-noise ratio. We show this effect in Fig.
\ref{fig_multi_emi}, where a mask-deconvolved image ("sky image")
is presented for a field including three sources. In the coding
direction the source position shows up as a sharp peak (PSF-wide),
while in the non-coding direction the true position corresponds to
a broad but still identifiable peak (as a function of the
\textbf{"zone"}, in the attitude correction algorithm). In the SA
data processing, a catalogued source identification is then done
by requesting that the SA measured position lies within $\pm$3
arcmin in the coding direction, and within $\pm$1 zone (i.e.,
$\pm$4$^{\circ}$) in the non coding direction.

The software package implementing this imaging procedure, named
EMI (Enhanced Multi sky Imaging), is applied to the
one-dimensional sky images obtained by co-adding data from
homologous detectors (i.e, the two pairs of detectors encoding the
X and Z directions) to detect, localize and identify the
significant sources in the field, independently in both
coordinates. The source flux is derived by integrating the peak
counts over a PSF width and then normalizing them to the net
source exposure (that is, excluding the times when the source was
occulted by the Earth or the satellite was passing through the
South Atlantic Anomaly) and to the effective area at the relevant
FOV position. The resulting normalized intensity, in units of
counts cm$^{-2}$ s$^{-1}$, is a measurement of the source flux and
it is then independent of the source position in the FOV.

The EMI procedure is able to detect and iteratively subtract the
source peaks, in order to identify weaker sources, but in doing
this it does not simulate the instrument response to the source
and it is then not suited for the subtraction of the coding noise
due to the partial coding of the detected sources (the whole
SuperAGILE field of view is indeed only partially coded, except
for the on-axis direction). To this purpose a dedicated IROS-like
(Iterative Removal Of Sources) procedure was developed, fitting
the complete imaging response to a detected source, in order to
iteratively subtract it from the image and search for the
subsequent, dimmer source. Compared to similar packages for past
experiments, the IROS procedure for SA must be able to account for
its several departures from ideality, that make its operation far
more complex than usual. The attitude variations of the AGILE
spacecraft amount to one SuperAGILE PSF every second
(0.1$^{\circ}$/s), within a range of $\sim$1$^{\circ}$ ($\sim$40
times the $\sim$1.5 arcminutes SA point source location accuracy,
see next section). The electronic noise in the front-end
electronics is large due to the high operative temperature (always
in the range $\sim$25-30$^{\circ}$C) and it is variable due to the
large temperature variations ($\sim$3-5$^{\circ}$C) along the
orbit. The electronic noise critically affects the imaging
capabilities for signal amplitudes near the low energy threshold
(see Pacciani et al. 2008a for an extensive discussion). All these
effects, also combined with the one-dimensional imaging capability
of the experiment, cause significant instabilities to the
simulation of the observed sky by the IROS procedure. Due to these
intrinsic difficulties, all the above experiment characteristics
had to be studied and understood from the flight data, and
modelled in the software. This data analysis package, named
EMIROS, is now near completion and it will be used to improve the
current imaging sensitivity of the experiment.

The analysis presented in this paper is based on the EMI data
processing only, and this implies that it has a limiting
sensitivity worse than the intrinsic experiment potential. The
coding noise of strong sources in the image in some cases affects
the flux estimate of the weaker sources, and this is taken into
account by adding a systematic uncertainty. This issue is
discussed in the next section.

\section{Status of Experiment Calibration}
\label{calib}

The SuperAGILE experiment was calibrated on ground at different
stages of its integration, from the "naked" silicon detectors up
to the full experiment inside the spacecraft in flight
configuration. Having a very good angular resolution (6 arcmin)
over a large field of view ($\sim$1 steradian), but a poor energy
resolution ($\sim$8 keV, full width at half maximum) over a narrow
bandpass (18-60 keV), SuperAGILE acts mostly as an X-ray imager.
For this reason, most of the calibration and software efforts were
devoted to its imaging properties and to the calibration of the
integrated effective area.

The wide field of view of the experiment was scanned with
radioactive sources of different energies during the ground
calibrations. Similarly, in orbit a series of 25 1-day pointings
at the beginning of the mission were devoted to observe the Crab
Nebula in different positions of the FOV. They were used to verify
and refine the results of the on-ground calibrations, in terms of
point source location accuracy and effective area, as a function
of the source position in the FOV. An onboard electronic circuit
allows to monitor and verify periodically the stability of the
gain and linearity of the experiment. A detailed description of
the on-ground and in-flight calibrations, their analysis, results
and estimation of the systematic effects may be found in
Donnarumma et al. (2006), Evangelista et al. (2006), and Feroci et
al. (2008). Here we summarize those results most relevant to the
data interpretation.

Due to the mentioned time/temperature-dependent threshold effects
combined with the poor energy resolution (see Pacciani et al.
2008a for details), the energy range between 17 and 20 keV is very
difficult to calibrate accurately, yet it is an important one, due
to the effects of the Silicon quantum efficiency and the energy
spectra of cosmic sources, both favored by a low energy response.
This energy range is then used only when the short-term
signal-to-noise ratio is high enough to permit to neglect the
threshold effects, on timescales shorter than the typical
temperature variations. This is usually achieved on
short-timescale transients, like gamma ray bursts or type I X-ray
bursts, while for longer integrations on the persistent emission
of steady sources we chose to limit our analysis to energies above
20 keV.

The point source location accuracy (PSLA) was verified in flight
with the detection of tens of known sources (see $\S$
\ref{detected_sources} and $\S$ \ref{grb})). By using hundreds of
detections in different positions of the field of view, the PSLA
was determined as 1.5 arcmin (90\% containment radius) for
detections with statistical significance higher than 10 $\sigma$.
A large systematic contribution ($\sim$1 arcmin) derives from the
knowledge of the true satellite attitude. In principle, one could
expect a contribution to the PSLA also by the angular extension of
the Crab Nebula, observed as $\sim$1-2 arcminutes below 10 keV.
The contribution of such an extension can be quantitatively
neglected when the 6-arcmin point spread function of SA is
considered. We verified that also the determination of the PSLA
provides consistent results both using the localizations of the
Crab Nebula only, and the localization of the other known sources
(see $\S$ \ref{detected_sources}), thus suggesting a low surface
brightness of the outer regions of the Crab Nebula with the SA
bandpass.

The source flux reconstruction is still significantly affected by
the systematics induced by a non perfect calibration of the
threshold and thermal-noise effects of the individual channels, as
well as by coding noise effects when a strong source (in terms of
detected counts) is in the field. Based on the mask design, the
amplitude of the coding noise induced by a source in other
positions of the FOV is expected to be at most 8\% of its peak
flux.

\begin{figure}
\centering
\includegraphics[width=9.4cm,clip]{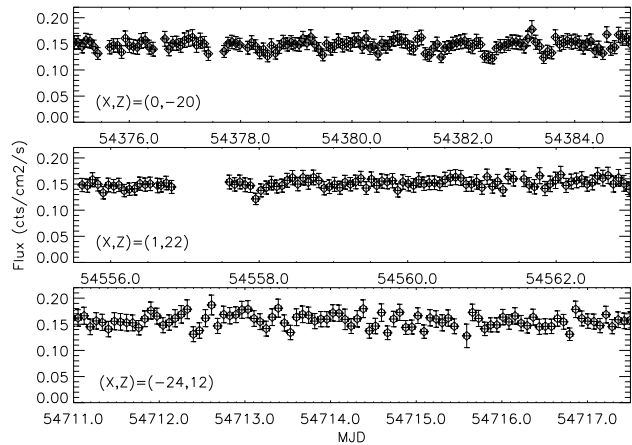}
\caption{Examples of $\sim$week-long sequences of automatic
orbital detection of the Crab for different positions in the SA
field of view. } \label{fig_crab_multi}
\end{figure}

Using the raster scan of the experiment field of view with the
Crab Nebula we were also able to study the systematics of the
source flux determination in the different positions of the FOV,
for the case of one only or dominant source in the field. We
selected positions in the SA field of view where the exposed area
is at least 20\% of the on-axis value on at least one of the two
directions (X or Z). Through the automatic orbital analysis, we
then used the Crab observations to estimate the systematic
uncertainty in the source flux determination. A set of $\sim$500
orbital detections of Crab were found to satisfy the above
criterium, and they provide a (statistically) constant Crab value
in the 20-60 keV energy range of 0.1511 counts cm$^{-2}$ s$^{-1}$,
provided that a systematic uncertainty equal to 8\% of the source
counts is added in quadrature to the statistical uncertainty of
the individual measurement. In Fig. \ref{fig_crab_multi} we show,
as examples, three $\sim$1-week stretches of orbital Crab
detections, in three different positions in the field of view,
demonstrating the source flux stability to within the quoted
uncertainties.

The calibration of the orbital determination of the source flux
was also successfully tested by comparing the simultaneous
observations of variable sources with SA and Swift/BAT, using the
publicly available BAT
lightcurves\footnote{http://swift.gsfc.nasa.gov/docs/swift/results/transients/}
in nearly the same energy range (BAT reports the 15-50 keV energy
range). The calibration of this type of data is particularly
relevant to the use of the SA orbital light curves of sources in
public distribution (see $\S$\ref{sa_web}).

\section{The AGILE Pointing Strategy and the SuperAGILE sky coverage}
\label{pointings}

The AGILE satellite has fixed solar panels and the spacecraft
attitude control system requires the solar panels oriented to the
Sun direction to within $\pm$1$^{\circ}$ at any time. Therefore,
the allowed range of pointing directions satisfying the Sun
constraint is included in a 2$^{\circ}$ wide circle, centered at
90$^{\circ}$ from the Sun, and it drifts by 1$^{\circ}$ per day in
order to keep the constraint always satisfied. This odd condition
is less critical for the AGILE scientific payload than it would be
in other cases, thanks to very large field of view of the GRID and
SA experiments.

From the scientific point of view, the primary experiment onboard
the AGILE mission is the GRID and the optimization of its
observing program is the driver for the pointing strategy of the
AGILE mission (e.g., Pittori et al. 2009). The effective field of
view of the GRID is significantly wider (by almost a factor of 2)
than that of SuperAGILE, posing looser pointing constraints to
reach a reasonably uniform coverage of the sky. As a result, the
integrated SuperAGILE exposure map after the period July 2007 -
April 2009 is largely non-uniform, and it mostly tracks the
average AGILE boresight position history. In Table
\ref{table:pointings} we list the sky regions at which the AGILE
mission pointed during the first $\sim$2 years. Updated AGILE
pointing information for both past and future observations may be
obtained at the web site of the AGILE Data Center
($http://agile.asdc.asi.it/$), part of the ASI Science Data Center
(ASDC).

\begin{table*}
\caption{A list of the main AGILE pointings during the first two
years. (*) Average coordinates.}
\label{table:pointings}      
\centering                          
\begin{tabular}{c c c c c}        
\hline\hline                 
Sky Region    & Date Start & Date Stop  & RA      & Dec  \\    
\hline                        
Vela Field          & 2007-07-13 & 2007-07-24 & 158.0 & -60.2 \\
                    & 2007-07-30 & 2007-08-01 & 150.8 & -70.2 \\
                    & 2007-08-02 & 2007-08-12 & 176.0 & -66.1 \\
                    & 2007-08-13 & 2007-08-22 & 195.6 & -66.6 \\
                    & 2007-08-23 & 2007-08-27 & 217.0 & -64.4 \\
                    & 2008-01-08 & 2008-02-01 & 147.0 & -62.5 \\
Galactic Anti-center & 2007-08-01 & 2007-08-02 & 37.1 & 12.7  \\
                    & 2007-08-12 & 2007-08-13 & 47.4 & 16.1  \\
                    & 2007-08-22 & 2007-08-23 & 57.1 & 18.6  \\
                    & 2007-09-01 & 2007-09-04 & 68.1 & 20.6 (*)  \\
                    & 2007-09-12 & 2007-09-13 & 78.5 & 21.7  \\
                    & 2007-09-15 & 2007-10-13 & 92.3 & 21.9 (*) \\
                    & 2007-10-22 & 2007-10-23 & 120.5 & 18.9  \\
                    & 2007-11-01 & 2007-11-02 & 130.6 & 16.3  \\
                    & 2008-03-30 & 2007-04-10 & 107.8 & 26.7 (*) \\
                    & 2009-03-31 & 2009-04-07 & 102.7 & 31.7 \\
Galactic Center     & 2007-08-27 & 2007-08-27 & 236.6 & -41.9  \\
                    & 2007-10-13 & 2007-10-22 & 290.9 & -18.9 \\
                    & 2007-10-23 & 2007-10-24 & 301.2 & -17.1  \\
                    & 2008-03-01 & 2008-03-30 & 254.7 & -39.8 \\
                    & 2008-09-10 & 2008-10-10 & 256.5 & -28.5 \\
                    & 2009-02-28 & 2009-03-25 & 247.0 & -29.0 \\
                    & 2009-03-25 & 2009-03-31 & 275.7 & -30.5 \\
Cygnus Field        & 2007-11-02 & 2007-12-01 & 296.9 &  34.5 \\
                    & 2007-12-01 & 2007-12-16 & 315.0 &  45.0 \\
                    & 2008-05-10 & 2008-06-09 & 304.3 &  36.0 \\
                    & 2008-06-15 & 2008-06-30 & 323.2 &  50.1 \\
                    & 2008-10-31 & 2008-11-30 & 295.5 &  35.6 \\
                    & 2008-11-30 & 2009-01-12 & 327.7 &  49.0 \\
                    & 2009-01-19 & 2009-02-28 & 325.7 &  68.1 \\
Musca Field         & 2008-02-14 & 2008-03-01 & 191.9 & -71.9 \\
                    & 2008-08-15 & 2008-08-31 & 175.3 & -74.1 \\
Aquila Field        & 2008-10-17 & 2008-10-31 & 291.0 &  10.1 \\
                    & 2009-04-07 & 2009-04-30 & 289.9 &  -1.6 \\
Antlia Field        & 2008-06-30 & 2008-07-25 & 161.8 & -47.7 \\
Vulpecula Field     & 2008-04-10 & 2008-04-30 & 286.3 &  20.8 \\
Virgo Region        & 2007-12-16 & 2008-01-08 & 173.4 &  -0.4 \\
"Field 8"           & 2007-09-04 & 2007-09-12 &  51.4 &  71.0 \\
                    & 2007-09-13 & 2007-09-15 &  74.9 &  58.3 \\
North Galactic Pole & 2008-04-30 & 2008-05-10 & 250.1 &  72.5 \\
South Galactic Pole & 2008-02-01 & 2008-02-09 &  58.3 & -37.8 \\
                    & 2008-02-12 & 2008-02-14 &  65.7 & -35.7 \\
ToO W Comae         & 2008-06-09 & 2008-06-15 & 182.3 &  29.6 \\
ToO S50716+714      & 2007-10-24 & 2007-11-01 & 155.5 &  67.4 (*)\\
ToO Mkn 421         & 2008-02-09 & 2008-02-12 & 251.0 &  50.3 \\
ToO 3C454.3         & 2007-07-24 & 2007-07-30 &  17.8 &  36.7 \\
                    & 2008-07-25 & 2008-08-15 &  22.3 &  39.1 \\
ToO SGR0501+4516    & 2008-08-31 & 2008-09-10 &  61.9 &  44.1 \\
ToO PKS0537-441     & 2008-10-10 & 2008-10-17 &  98.8 & -46.8 \\
ToO Carina Field    & 2009-01-12 & 2009-01-19 & 161.7 & -59.9 \\

\hline                                   
\end{tabular}
\end{table*}

\begin{figure}
\includegraphics[width=9cm,angle=90,clip]{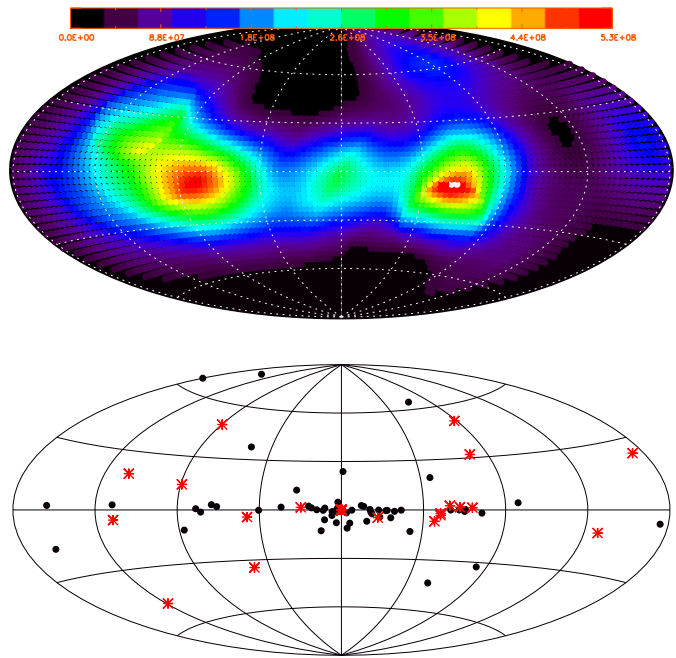}
\caption{\textit{Top Panel}: Integrated exposure map, in units of
effective area times exposure time, to any region of the sky, over
the time period July 2007 - April 2009. \textit{Bottom Panel}:
Distribution of the sources detected by SuperAGILE in galactic
coordinates. Red asterisks show the sky distribution of the
localized gamma-ray bursts.} \label{fig_exposure}
\end{figure}

Taking into account the SA effective area as a function of the
off-axis angle, we can then derive the exposure map, that is the
cumulative distribution of the product of the net exposure time
and the effective area, in units of cm$^{2}$ s, as shown in Figure
\ref{fig_exposure} (top panel). The false-color  exposure map in
Galactic coordinates clearly shows the effect of the AGILE long
exposures to the Vela and Cygnus regions, that are favored not
only by the presence of several scientific and calibration targets
for the GRID experiment, but also by the Sun constraints.
Relatively deep exposures were also obtained to the Galactic
Center and anti-Center regions (the latter due to the presence of
the Crab Nebula), and to the Virgo region, where a coordinated
multi-frequency campaign was spent (see Pacciani et al., 2009).

\section{Science with SuperAGILE}
\label{science}

The characteristics of the SuperAGILE experiment make it suited
for the detection of bright transients and the extended monitoring
of bright Galactic sources. The one-dimensional imaging, the
spacecraft attitude control system and constraints make it not
well suited for very long integrations (i.e., longer than $\sim$1
week). However, thanks to its large field of view, SuperAGILE is
able to provide month-long continuous light curves for several
sources at the same time, with bin sizes from minutes to
$\sim$days, depending on the source intensity and position in the
field of view. For bright sources timing and spectral analysis can
also be carried out using the photon-by-photon data. In addition
to the persistent sources, SA typically detects and localizes one
gamma ray burst per month, as well as other Galactic or
extragalactic transients.

In the following sections we report about the sources for which
flux measurements with SA are available so far and provide some
details about specific noticeable regions in the sky, as
case-studies useful to illustrate the results of both automated
and manual analysis of the SA data. Deeper analysis and
interpretation for a few specific sources is the subject of
separate papers, as in the case of Mkn 421 (Donnarumma et al.
2009a) or 3C 273 (Pacciani et al. 2009), as well as the gamma-ray
bursts (Del Monte et al. 2008a).

\subsection{Detected sources}
\label{detected_sources}

During its first $\sim$20 months of observations, obtained between
mid-July 2007 and end-April 2009 with the satellite $\sim$94\% of
time in nominal pointing and the experiment in nominal
configuration, SuperAGILE observed the sky with significantly
different exposure to various regions. The typical pointing
strategy of the AGILE mission is to have month-long observations
of the same field, with the Sun constraints imposing a drift of
the observed field by approximately 1$^{\circ}$ per day. During
this period SuperAGILE was able to monitor a few tens of sources,
typically brighter than $\sim$30-40 mCrab.

Table \ref{table:list} summarizes the current status of the SA
source detections. The results reported in the table derive mostly
from the automated daily integrations (i.e., $\sim$40ks net
exposure each). In some other cases, specified in the "Timescale"
column, the reported source was detected through its transient
emission, measured by means of a dedicated analysis (e.g.,
Galactic transients). For each source it is specified whether it
was also detected by the automated analysis on the orbital
timescale, implying that flux measurements for it are publicly
available at the ASDC website. We note that the Table
\ref{table:list} is not intended as a catalog. In fact, the
analysis that provided such results was not carried out
homogeneously and systematically over the full data archive, and
it was not optimized for the source detection (see Sect.
\ref{software}). In addition, shorter and longer integration times
were not systematically explored. A complete re-analysis of the
full SA data archive is planned with the final EMIROS software
package. A complete SA source catalog will be then subject of a
future publication, where a more homogeneous and complete analysis
of the SA data archive will be reported.

The columns in the table report, for each source, after the name:
the celestial coordinates, the average flux measured by SA in the
20-60 keV, the total exposure time, the maximum significance of
the individual SA measurements (typically on daily timescale), the
source type, the availability of orbital measurement data, and a
column for any notes required by that specific source. It is
important to note that the Average Flux column in the table
provides the source flux measurements obtained by positive
detections only (i.e., above 4$\sigma$ within a region of the SA
field of view exposing no less than 20\% of the on-axis effective
area, even if in only one of the two SA coordinates). As such,
these numbers are biased upward. Therefore, these flux
measurements are not to be taken as an average flux level of
individual sources over the 2007-2009 time interval. Similarly, we
note that the results of the automated orbital analysis (see
$\S$\ref{sa_web}), also reporting positive detections only, are
even more biased upward, due to the lower sensitivity of the
experiment on such a shorter time scale. Thus, the average flux
measurement in the orbital analysis is expected to be
systematically higher than in the daily integrations (except for
bright sources), because it selects the times when the source
shows a flux higher than the SA orbital sensitivity. The exposure
time column is meant to provide the reader with a general
information on how long a specific source has been at a flux level
detectable by SuperAGILE during its visibility periods. It is
computed by adding up the integrations when the source was
positively detected, filtering out the time when the source was
occulted by the Earth and when the satellite was passing through
the South Atlantic Anomaly.

%
\begin{table*}
\caption{The list of X-ray sources detected by SuperAGILE in the
period July 2007 - April 2009. Average Flux (mCrab, 20-60 keV),
Total Exposure and Maximum detection significance are based on
daily averages. A "Y" in the "Orbital Data" column specifies that
the source was also detected on the orbital timescale, while the
Note column identifies specific integrations (e.g., short
transients). The source type is identified according to the
following classes: HMXB - High Mass X-ray Binary; LMXB - Low Mass
X-ray Binary; SGR - Soft Gamma Repeater; PWN - Pulsar Wind Nebula;
Sy - Seyfert Galaxy; BL - BL Lac; RG - Radio Galaxy; AXP -
Anomalous X-ray Pulsar. See text for other details. }
\label{table:list}      
\centering                          
\begin{tabular}{c c c c c c c c c c}        
\hline\hline                 
Name\footnotemark[1] & RA    & Dec   & Average Detected Flux & Total Exposure & Max. Sign. & Type & Orbital & Note  \\    
     & (deg) & (deg) & (mCrab)               & (ks)     & ($\sigma$) &      & Data    &    \\    
\hline                        

\hline
\\
          SMC X-1 &  19.271 & -73.443 &   87 & 140  &   4.4 &         HMXB &    N &   \\
      1A 0114+650 &  19.511 &  65.292 &   13 & 380  &   5.9 &         HMXB &    N & 5-h flare \\
            X Per &  58.846 &  31.045 &   55 &  45  &   4.5 &         HMXB &    Y &   \\
    SGR 0501+4516 &  75.278 &  45.276 &   -  & -    &    -  &          SGR &    N & Short bursts \\
          LMC X-4 &  83.207 & -66.370 &   50 & 400  &  10.1 &         HMXB &    N &   \\
             Crab &  83.628 &  22.020 & 1000 & 2477 &  76.4 &          PWN &    Y &   \\
         Vela X-1 & 135.512 & -40.557 &  603 & 1523 &  47.6 &         HMXB &    Y &   \\
     GRO J1008-57 & 152.441 & -58.291 &   56 & 220  &   6.0 &         HMXB &    N &   \\
          Mkn 421 & 166.114 &  38.209 &   40 & 140  &   9.4 &           BL &    N &   \\
       1A 1118-61 & 170.238 & -61.917 &  475 & 447  &  47.6 &         HMXB &    Y &   \\
          Cen X-3 & 170.300 & -60.638 &  173 & 1859 &  21.3 &         HMXB &    Y &   \\
   1E 1145.1-6141 & 176.869 & -61.953 &   44 & 1010 &   9.4 &         HMXB &    Y &   \\
         NGC 4151 & 182.636 &  39.405 &   17 &  115 &   6.9 &           Sy &    N &   \\
         GX 301-2 & 186.657 & -62.770 &  390 & 3315 & 129.9 &         HMXB &    Y &   \\
           3C 273 & 187.278 &   2.052 &   24 &  740 &  21.3 &          QSO &    N &   \\
            Cen A & 201.365 & -43.019 &   60 & 200  &   6.6 &           RG &    Y &   \\
   1E1547.0-5408  & 237.725 & -54.307 &    - &   -  &     - &          AXP &    N & Short bursts \\
      4U 1608-522 & 243.179 & -52.423 &  126 & 130  &  8.22 &         LMXB &    Y &   \\
          Sco X-1 & 244.979 & -15.640 & 1666 & 3594 & 103.0 &         LMXB &    Y &   \\
      4U 1624-490 & 247.011 & -49.198 &  139 &  8   &   4.6 &         LMXB &    N &   \\
  IGR J16318-4848 & 247.966 & -48.808 &  165 &  360 &  22.6 &         HMXB &    Y &   \\
       4U 1626-67 & 248.070 & -67.461 &   53 & 2600 &  11.7 &         LMXB &    Y &   \\
      4U 1636-536 & 250.231 & -53.751 &   68 &  380 &   8.5 &         LMXB &    Y &   \\
         GX 340+0 & 251.448 & -45.611 &   85 &  908 &  18.5 &         LMXB &    Y &   \\
          Her X-1 & 254.458 &  35.342 &  200 &  122 &   5.3 &         HMXB &    N &   \\
     OAO 1657-415 & 255.199 & -41.673 &   96 &  530 &  27.1 &         HMXB &    Y &   \\
         GX 339-4 & 255.706 & -48.789 &  105 &  868 &  11.3 &         LMXB &    Y &   \\
      4U 1700-377 & 255.986 & -37.844 &  357 & 1179 &  35.8 &         HMXB &    Y &   \\
         GX 349+2 & 256.435 & -36.423 &  117 & 3110 &  21.4 &         LMXB &    Y &   \\
      4U 1702-429 & 256.563 & -43.035 &   63 &  561 &  10.8 &         LMXB &    Y &   \\
      4U 1705-440 & 257.226 & -44.102 &   78 &  805 &  14.5 &         LMXB &    Y &   \\
         GX 354-0 & 262.989 & -33.834 &   61 & 1977 &  18.6 &         LMXB &    Y &   \\
           GX 1+4 & 263.009 & -24.745 &   64 &  805 &   7.8 &         LMXB &    Y &   \\
      4U 1735-444 & 264.742 & -44.450 &   47 &  272 &  11.8 &         LMXB &    N &   \\
    1E1740.7-2942 & 265.978 & -29.745 &   47 & 1251 &  11.2 &         LMXB &    Y &   \\
  IGR J17464-3213 & 266.565 & -32.233 &  134 &  536 &  23.1 &         LMXB &    Y &   \\
  IGR J17473-2721 & 266.825 & -27.344 &   37 &  129 &   5.3 &         LMXB &    N & X-ray burst  \\
           GX 3+1 & 266.983 & -26.563 &   40 &  276 &  10.0 &         LMXB &    Y &   \\
  AX J1749.1-2639 & 267.300 & -26.647 &  196 & 1097 &  37.2 &         HMXB &    Y &   \\
 SAX J1750.8-2900 & 267.600 & -29.038 &   67 &  137 &  21.8 &         LMXB &    N & X-ray burst \\
SWIFTJ1753.5-0127 & 268.368 &  -1.453 &  116 &  158 &   6.0 &         LMXB &    N &   \\
           GX 5-1 & 270.284 & -25.079 &  137 & 2621 &  33.4 &         LMXB &    Y &   \\
     GRS 1758-258 & 270.301 & -25.743 &  100 &  791 &  28.2 &         LMXB &    N &   \\
           GX 9+1 & 270.385 & -20.529 &   50 & 1260 &   8.4 &         LMXB &    N &   \\
 SAX J1808.4-3658 & 272.115 & -36.979 &    - &    - &     - &         LMXB &    N & X-ray burst \\
      SGR 1806-20 & 272.164 & -20.411 &    - &  -   &     - &          SGR &    N & Short bursts \\
 SAX J1810.8-2609 & 272.685 & -26.150 &   35 &  376 &   8.4 &         LMXB &    Y &  \\
          GX 13+1 & 273.631 & -17.157 &   46 &  371 &   5.5 &         LMXB &    Y &   \\
      1M 1812-121 & 273.800 & -12.083 &   45 &  708 &   7.8 &         LMXB &    Y &   \\
          GX 17+2 & 274.006 & -14.036 &  122 & 2705 &  22.4 &         LMXB &    Y &   \\
      4U 1820-303 & 275.919 & -30.361 &   80 & 3099 &  26.6 &         LMXB &    Y &   \\
      3A 1822-371 & 276.445 & -37.105 &   53 & 1016 &   7.9 &         LMXB &    Y &   \\
    Ginga 1826-24 & 277.367 & -23.796 &   84 & 3316 &  17.3 &         LMXB &    Y &   \\
   Ginga 1843+009 & 281.412 &   0.891 &   95 &  125 &   5.2 &         HMXB &    N &   \\
HETE J1900.1-2455 & 285.036 & -24.920 &   60 &  210 &   7.6 &         LMXB &    Y &   \\
     GRS 1915+105 & 288.798 &  10.946 &  347 & 4361 &  66.8 &         LMXB &    Y &   \\
       4U 1954+31 & 298.926 &  32.097 &   64 &  279 &  13.8 &         LMXB &    Y &   \\
          Cyg X-1 & 299.590 &  35.202 &  714 & 7923 & 161.4 &         HMXB &    Y &   \\
     EXO 2030+375 & 308.063 &  37.638 &   92 & 1976 &  28.7 &         HMXB &    Y &   \\
          Cyg X-3 & 308.107 &  40.958 &  165 & 6290 &  32.8 &         HMXB &    Y &   \\
          Cyg X-2 & 326.172 &  38.322 &   59 & 3793 &  16.1 &         LMXB &    Y &   \\

\\

\hline                                   
\end{tabular}

\end{table*}

In Fig. \ref{fig_exposure} (bottom panel) we show the sky
distribution, in galactic coordinates, of the sources detected by
SA and listed in Table \ref{table:list}. As expected, the sources
are mostly located in the Galactic center and where AGILE spent
the largest fraction of its observing time, the Vela and Cygnus
regions.

\subsection{The Vela region}
\label{vela}

In this and in the next section we show analysis of two
significant fields observed by SuperAGILE, representing conditions
with a few and many sources, the Vela region and the Galactic
Center. We use them as case studies to describe and explain the
typical results of the scientific analysis of the SuperAGILE data
in two different conditions.

\begin{figure}
\centering
\includegraphics[width=8.4cm,clip]{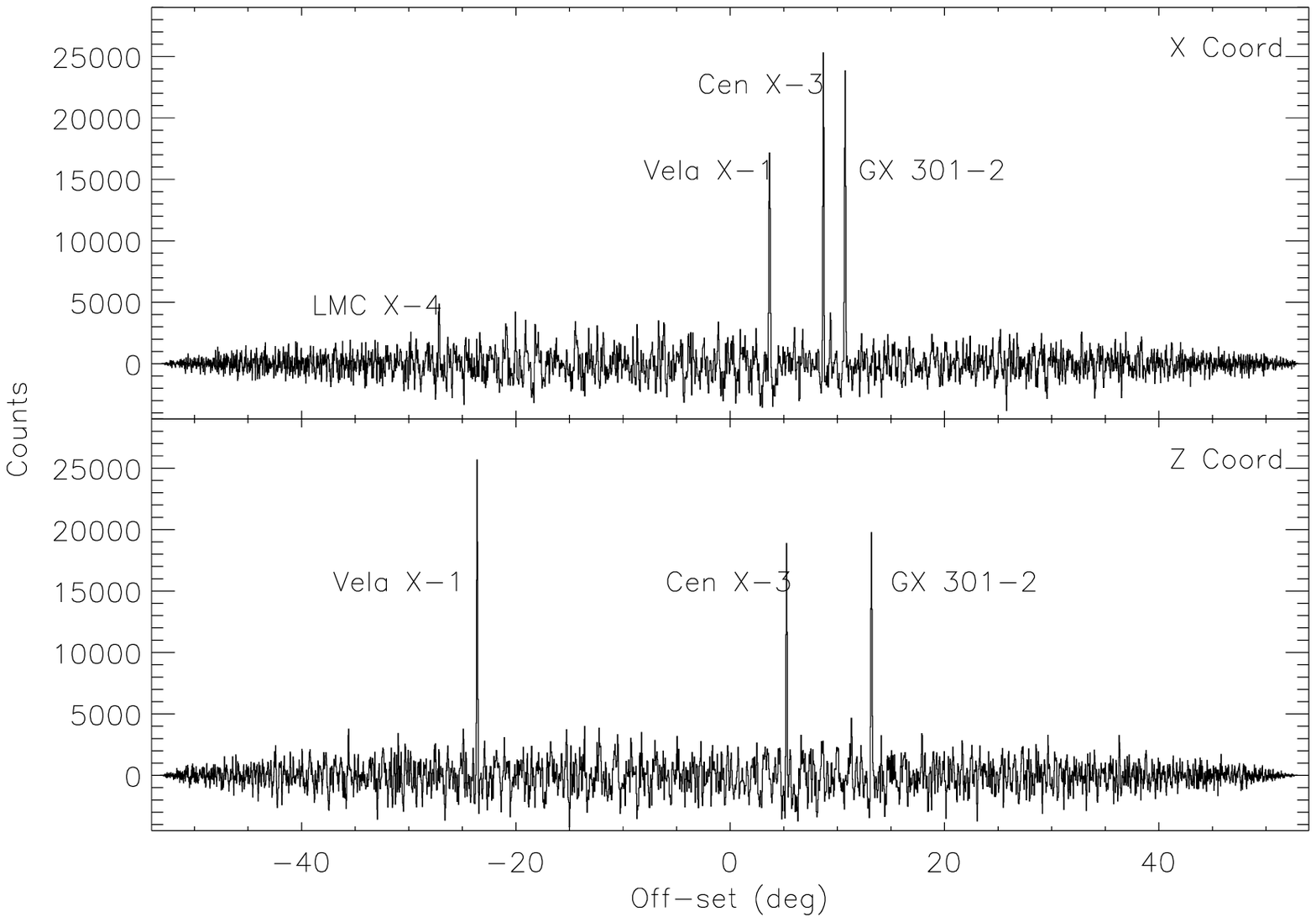}
\caption{One day exposure to the Vela field. Data refer to MJD
54475 (10th January 2008). Center of the field of view (X,Z)=(0,0)
is at RA=149.31, Dec=-62.95 (J2000), corresponding to galactic
coordinates l=284.55, b=-6.48.} \label{fig_vela_field}
\end{figure}

In Fig. \ref{fig_vela_field} we provide the set of two one
dimensional sky images obtained by a one-day integration centered
at RA=149.31, Dec=-62.95 (J2000, corresponding to galactic
coordinates l=284.55, b=-6.48) on 10th January 2008. The abscissae
report the angular off-set with respect to the center of the SA
field of view, in the X and Z coordinates of the SA reference
frame (+Z aims at the Sun, +Y is the pointing direction). The bin
size is the SA pixel, corresponding to 3 arcminutes at the center
of the field of view. The ordinates show the counts per bin for
the sum of the two detectors encoding the same direction (X or Z),
after a deconvolution with the mask code. With reference to the
"two dimensional encoding" caused by the spacecraft attitude
variations (see $\S$ \ref{software}), the most appropriate
representation of a SA sky image, for each of the X and Z
coordinates, would be two-dimensional, similar to Fig.
\ref{fig_multi_emi}, where the proper attitude correction is
applied to every region in the field of view. However, for the
sake of clarity we chose to have one dimensional images,
displaying the on-axis non-coding zone for every sky bin (i.e.,
X=0 for the Z image, and vice versa), except for those where a
source is detected, for which the relevant non-coding zone
correction is shown, in order to image the detected source
properly. Thus, these one-dimensional images of the sky are a sort
of mosaic of different non-coding zones.

The ordinates of these images report the integrated counts, for
each bin. According to the shape of the SA PSF, in most of the
cases the source peak in the counts image corresponds to a
fraction varying between 70\% and 100\% of the total source counts
(see Evangelista et al. 2006 for details). The estimation of the
total source counts needs then to be done by proper integration
over the PSF shape at the position of the source.

It is useful to notice how the amplitude of the fluctuations in
Fig. \ref{fig_vela_field} decreases as the off-axis angle
increases. This is due to the fact that the coded imaging
subtracts the background counts, but not their fluctuations. They
depend on the total counts (background plus sources) in the
fraction of the detector exposed to each specific off-axis angle.
In other words, the on axis direction is seen by the full area of
the detectors, while increasing off-axis directions are seen with
decreasing fractions of it. This implies a smaller number of
counts in the involved detector region, and then smaller values of
their fluctuations. Consequently, in order to infer the
statistical significance of any peak in the image, the counts in
the peak must be compared to the value of the background
fluctuations (e.g., $\sigma$) at that specific position in the
field of view.

The one-day integration on the Vela field taken on MJD 54475 shows
the detection of a few sources in both X and Z SA coordinates,
noticeably Vela X-1, Cen X-3 and GX 301-2, here with an estimated
average flux of about 380, 130 and 195 mCrab, respectively. The
detection significance is 14, 22 and 21 $\sigma$ on the X
coordinate image, and 27, 17 and 19 $\sigma$ on the Z coordinate,
respectively. {The net exposure to each of the sources in this
image is 46, 56 and 59 ks, respectively. It is worth noticing the
wide angular separation between the sources, up to
$\sim$40$^{\circ}$ .

When looking at the SA images of the sky it is important to remind
how the angular response of the experiment works (see Feroci et
al. 2007 for details). The SA FOV is defined by a collimator that
causes a linear decrease of the effective area from the center of
the field to the edge. The shape of the FOV is rectangular (see
Fig.\ref{fig_fov}) and it is narrower in the non-coding direction,
implying that the effective area decreases more rapidly if a
source "moves" from the center to the edge of the FOV along the
non-coding direction than along the coding one. Since the peaks in
the SA images are in counts, the shape of the collimator response
makes the peak value for a given source strongly dependent on its
position in the FOV. This effect explains why the same two
sources, observed simultaneously, may show peaks of very different
relative heights in the X and Z images, being dependent on their
relative positions in the X and Z SuperAGILE coordinates. An
example of this effect is given by Cen X-3 and Vela X-1 in Fig.
\ref{fig_vela_field}: the peak of Vela X-1 is higher in the Z
image than in the X because in the latter it lies at a position
with smaller effective area. The reverse applies to Cen X-3. This
SA property also clarifies why the source LMC X-4 is significantly
detected only in the X image. The source is at an expected
position in the SA reference frame of (X,Z)=(-27.14, -0.87)
degrees. At this position the exposed area in the Z detectors (for
which the non-coding position is $\sim$27$^{\circ}$ off-axis) is
insufficient for a source detection, while in the X detectors it
is detected at 10 $\sigma$, with a flux of 70 mCrab.

Given the high detection significance of the sources in the daily
image integration, light curves can be obtained with finer time
resolution. In Fig. \ref{fig_vela_gx_cen} we show the light curve
of the same sources, obtained by the automated orbital analysis,
over a time span including MJD 54475, with a bin size
corresponding to the SA orbital time scale ($\sim$6 ks elapsed
time). The plot shows the average flux measurement obtained by
SuperAGILE at each AGILE orbit, with a net exposure to each of the
sources of about $\sim$3-4000 s. We note that, at variance with
the images, in the light curves the source counts are normalized
and directly represent flux values (under the assumption of a
Crab-like energy spectrum). When the source is not detected the
3-$\sigma$ upper limit at the position of the source is provided.
It is interesting to note, for example, how the $\sim$half-day
binary eclipse is nicely detected in the light curve of Cen X-3,
every $\sim$2.1 days.

\begin{figure}
\includegraphics[width=9cm,height=9cm,clip]{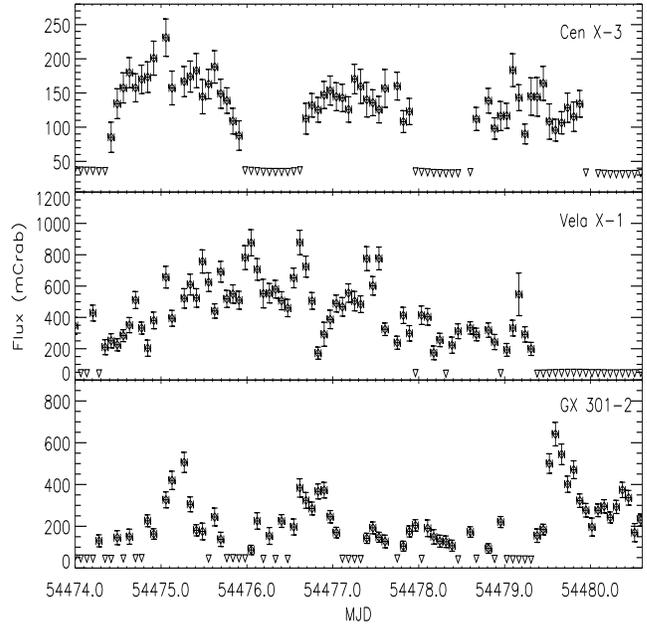}
\caption{SuperAGILE simultaneous orbital light curves for Vela
X-1, Cen X-3 and GX 301-2 between MJD 54474 and MJD 54480. Each
time bin corresponds to one AGILE orbit (approximately 100 minutes
real time). Upside-down triangles identify 3-$\sigma$ upper limits
when there was no detection. Vertical error bars are 1-$\sigma$
uncertainties, and include systematics. } \label{fig_vela_gx_cen}
\end{figure}

Thanks to the photon-by-photon data always transmitted to the
ground (see $\S$\ref{software}), a more detailed analysis can be
carried out on individual sources, provided the data have good
enough statistical quality. As an example, for the three X-ray
binary pulsars in the Vela field for which we provided images and
light curves above, SuperAGILE was able to detect their spin
periodicity and obtain their pulse profile by period folding. In
Fig. \ref{folding_vela_gx_cen} we show the folded light curves at
the periods $P \simeq$ 283 s for Vela X-1, $P \simeq$ 680 s for GX
301-2 and $P \simeq$ 4.8 s for Cen X-3 (the folding epoch is
arbitrary).

\begin{figure}
\includegraphics[width=9cm,clip]{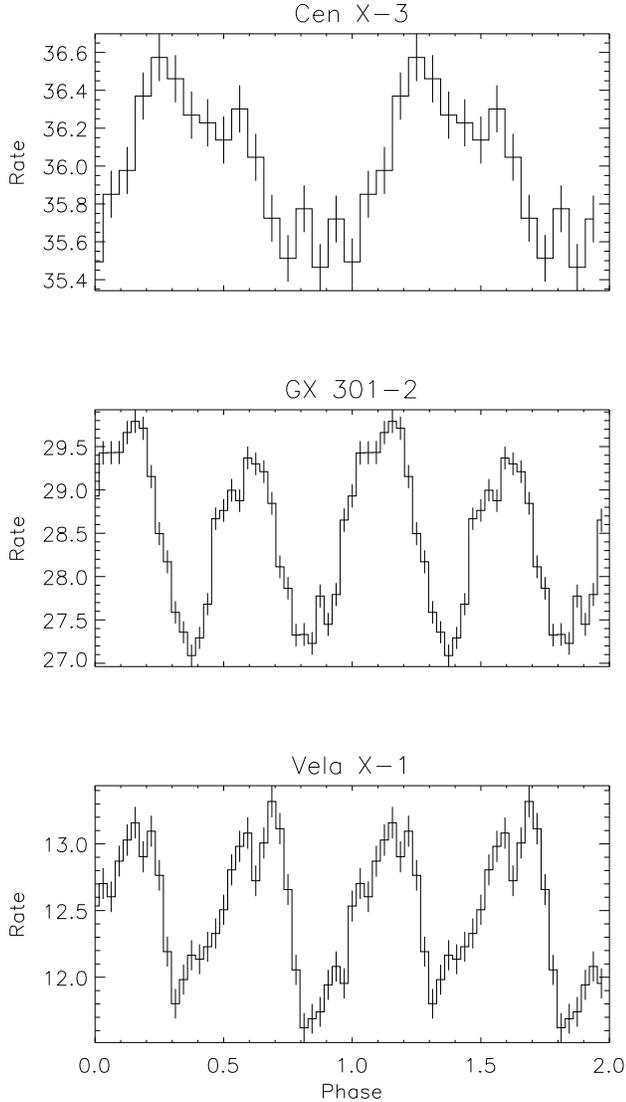}
\caption{SuperAGILE folded light curves for the X-ray pulsars Vela
X-1, Cen X-3 and GX 301-2. Data used here are from MJD 54753 for
Vela X-1, 54317 for Cen X-3 and GX 301-2. }
\label{folding_vela_gx_cen}
\end{figure}

\subsection{The Galactic Center}
\label{gc}

The Galactic Center region is the field most rich of sources for
any X-ray experiment. A number of bright sources are concentrated
within the central $\sim$10-20$^{\circ}$ of our Galaxy, mostly
neutron star and black-hole X-ray binaries, and every look at this
region offers a different picture, due to their large variability.
For SuperAGILE this field is also very complex, due to the
one-dimensional encoding of the sky, leading to source confusion
in some conditions. This is made even more difficult due to the
large field of view of SA, that often includes Sco X-1, the
brightest X-ray source in the sky, despite its $\sim23^{\circ}$
galactic latitude. Images taken by SA from this region of the
Galaxy need to be analyzed very carefully, especially when Sco X-1
is in the field of view, due to the higher fluctuations and coding
noise it induces.

The Galactic Center region was observed 5 times in 2007-2008 (see
Table \ref{table:pointings}): two short observations in 2007
(August and October), two one-month pointings in 2008 (March and
September) and in March 2009. In Fig. \ref{fig_gc} we show a
SuperAGILE image taken on October 2007 of a field centered near to
the Galactic Center (boresight at l=21.430, b=-20.031) such that
the position of Sco X-1 is at SA coordinates X=-6.252, Z=48.792,
that is outside the field of view of the X detectors (for these
detectors it has a non-coding off-set angle of $\sim$49$^{\circ}$,
and their effective area goes to zero at $\sim$35$^{\circ}$
off-set)), and marginal in Z. The favorable condition of not
having very bright sources in the field allowed for a 9-day long
integration, corresponding to typical net exposure to the sources
of about 370 ks. The X image (Fig. \ref{fig_gc}, top panel) is
nearly parallel to the Galactic longitude (i.e., the Galactic
plane is seen "face-on"), and it shows several sources, as
labelled in the figure: GRS 1915+105, 1M 1812-121, GX 17+2, GX
13+1, Ginga 1826-24, GX 5-1, SAX J1810.8-2609, 4U 1820-303, 3A
1822-371, and AX J1846.4-0258 and/or XTE J1855-026 (confused,
being projected to X coordinates separated by less than 0.4
arcmin). These sources are all X-ray binaries in our Galaxy, and
their intensity in this observation ranges from $\sim$40 to
$\sim$150 mCrab.

\begin{figure*}
\centering
\includegraphics[width=16.4cm,clip]{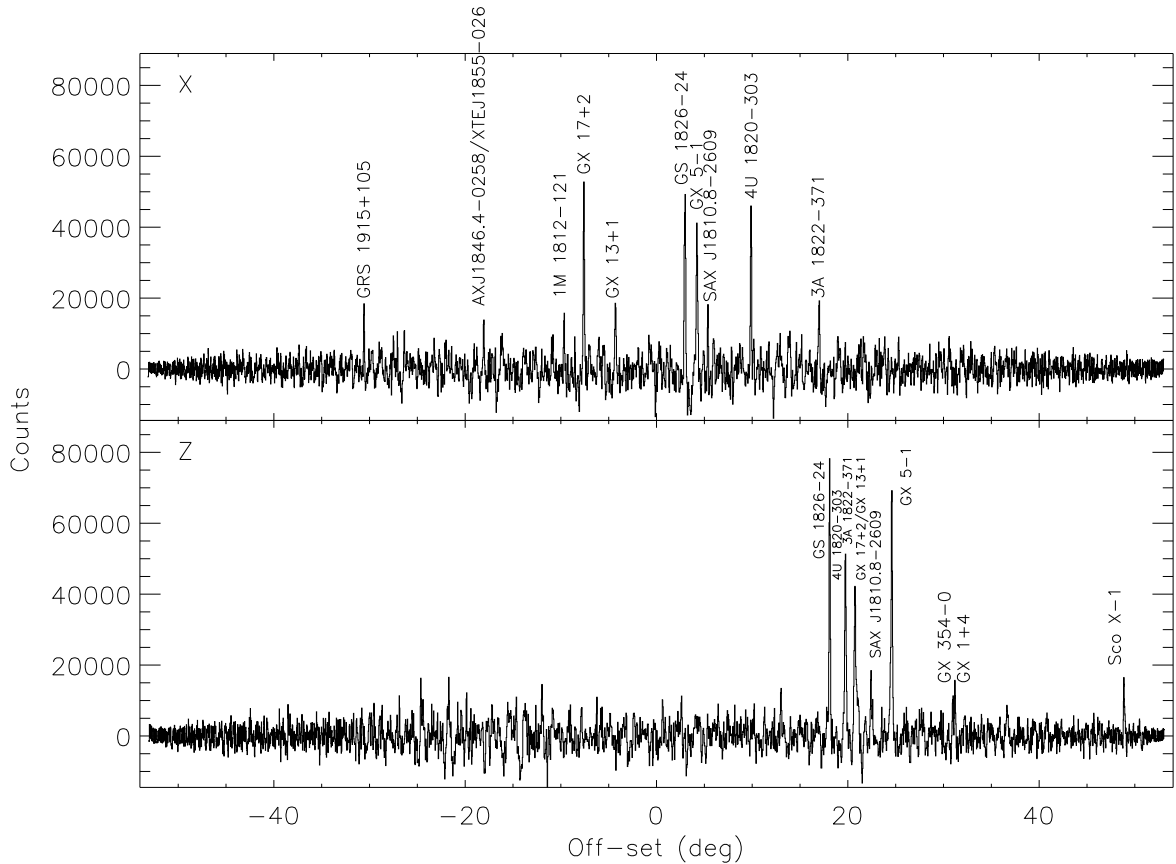}
\caption{A 600 ks long exposure to a field near to the Galactic
Center, from 13th to 22nd October 2007. The field is centered at
RA=296.265, Dec=-18.840 (l=21.430, b=-20.031). In these images,
the Galactic Center (l=0, b=0) is at SuperAGILE coordinates X=8.7,
Z=28.1. } \label{fig_gc}
\end{figure*}

The sources in the X image are approximately evenly distributed
over the field of view, thus the balance of counts does not give
rise to significant (concentrated) coding noise disturbance.
Instead, the Galactic plane projection onto the SA Z coordinate is
almost "edge-on" and the sources are clustered around an off-set
of $\sim$20-30 degrees (the Galactic Center is at Z $\simeq$
28$^{\circ}$). This fact, together with the bright Sco X-1 lying
about 49$^{\circ}$ off-axis, induces significant fluctuations due
to coding noise at the opposite side of the field of view, near
off-set -20$^{\circ}$. The sources that we identified in the Z
image are: Ginga 1826-24, 4U 1820-303, 3A 1822-371, GX 17+2, GX
13+1, SAX J1810.8-2609, GX 5-1, GX 354-0, GX 1+4 and Sco X-1.
These sources are not all those detected in the X image, due to
their different projections on the two SA coordinates. As an
example the microquasar GRS 1915+105 lies at SA coordinates
X=-30.607, Z=2.594, implying that it has a reasonable exposed area
in X detectors (being only 2.6$^{\circ}$ off-axis in their
non-coding direction) and a negligible area in the Z detectors
(for which it lies at 30.6$^{\circ}$ in the non-coding direction).
Thus, in the Z image we expect a far less significant peak for
this source, that is indeed detected in the expected position but
with a significance below our threshold. Similarly, the two
sources confused in the X image, at X=-18.03, AX J1846.4-0258 and
XTE J1855-026, are expected at Z=12.293 and Z=9.873 respectively.
Although their coordinates do not favor a detection in the Z
image, the presence of a small peak at Z=12.3 and nothing at Z=9.9
supports the identification with AX J1846.4-0258. Other weaker
sources are most likely detected in these images, but they will
significantly detected only after the coding noise due to the
brighter sources will be subtracted with the future version of the
SA data analysis software.

\subsection{Galactic transients} \label{galactic_transients}

In addition to the long term monitoring of variable sources, the
wide field of view and sensitivity of SuperAGILE allow for the
detection and localization of short timescale events, originating
from Galactic transient sources. This is the case for the short
bursts from Soft Gamma Repeaters (SGRs), with typical duration in
the range of few hundreds of ms, the type I X-ray bursts from Low
Mass X-ray Binaries (LMXB), in the range of tens of seconds, and
the outbursts or flares from High Mass X-ray Binaries (HMXB),
usually lasting from hours to days.

The relatively small fraction of time spent by AGILE pointing at
the Galactic Center region (see $\S$\ref{pointings} and Table
\ref{table:pointings}), where these transients are more frequent,
made the SA detection rather sparse. Type I X-ray bursts, having
thermal energy spectra with temperatures of about $kT\sim$2-3 keV,
are also unfavored by the band pass of SA. Type I X-ray bursts
were detected by SA from IGR J17473-2721, SAX J1750.8-2900, and
the millisecond pulsar SAX J1808.4-3658 (Del Monte et al. 2008b;
Pacciani et al. 2008b; Del Monte et al. 2008c). The case of IGR
J17473-2721 is noticeable because the single X-ray burst detected
by SA was the first one ever detected from this previously
unidentified source and allowed its classification as a LMXB. This
was confirmed by the later activity of the source detected by
other experiments: the complete outburst lasted for approximately
3 months, showed several bursts and an hard-to-soft spectral state
change followed by a transition back to a hard spectral state.
During the soft spectral state 900 Hz QPOs were also detected with
the RossiXTE/PCA, as typical of atoll sources (Altamirano et al.
2008).

Longer outbursts were detected from several X-ray binary systems:
the Compton-thick X-ray binary source IGR J16318-4848 (Pacciani et
al. 2008c), the Be binary GRO J1008-57 (Evangelista et al. 2008),
the HMXB 1A 1118-61, the slow spinning HMXB source 3A 0114+650
(Pacciani et al. 2009b), the neutron star transient SAX
J1750.8-2900 (Pacciani et al. 2008d, see Fig. \ref{saxj1750_lc}).
The recurrent transient and black hole candidate IGR J17464-3213
was detected by SA on September 2008. An interesting state of Vela
X-1 was also detected by SA on October 2008 (Soffitta et al.
2008): the source flux rose from an average 500 mCrab level up to
more than 2 Crab in approximately 7 hours, before redescending to
a 500 mCrab flux. The intensity of the source during the flare was
such that the individual 283.5-s pulsations of the neutron star
were clearly visible on the raw 16-s SA ratemeters counters.
Interestingly, during the hours-long flare, shorter $\sim$100 s
flares were also observed, peaking at fluxes as high as 4 Crabs.

\begin{figure}
\includegraphics[width=9cm,clip]{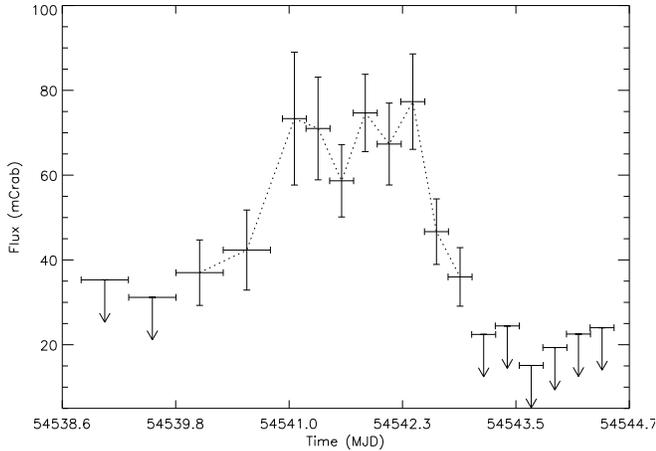}
\caption{A $\sim$1 day flare from the neutron star transient SAX
J1750.8-2900 detected by SuperAGILE on March 16-17, 2008 (Pacciani
et al. 2008d). Arrows show 3$\sigma$ upper limits, when no
significant detection is available.} \label{saxj1750_lc}
\end{figure}

During the last two years several magnetar sources entered new
periods of intense bursting activity, namely the SGRs 1806-20,
1627-41 and 0501+4516 (actually discovered by Swift/BAT on 22nd
August 2008, Barthelmy et al. 2008), as well as the Anomalous
X-ray Pulsar (AXP) 1E 1547.0-5408. Unfortunately, the periods of
activity of these sources generally did not coincide with the
observation of their sky regions by AGILE. Thus, only a handful of
short bursts were imaged by SA, originating from SGR 1806-20, 1E
1547.0-5408 and SGR 0501+4516. The latter was indeed monitored
during a dedicated target of opportunity observation approximately
10 days after its discovery, from 31st August to 10th September,
but the source turned out to have gone back to an almost
burst-quiescent state (see, e.g., Rea et al. 2009) and only 2
short bursts were detected by SuperAGILE (Feroci et al. 2008b).
Examples of short bursts from the SGR sources detected by SA are
shown in Fig. \ref{sgr_lc}, highlighting the time resolution
achieved by the photon by photon data, basically limited by the
counts statistics only. Sometimes the short bursts from these
active magnetars were so bright that they were detected in the
SuperAGILE rates even with the source lying well outside the
experiment field of view, passing through the collimator walls or
the AGILE satellite structures. This happened for a few events
from SGR 1806-20, SGR 0501+4516 and 1E1547.0-5408.

\begin{figure}
\includegraphics[width=4cm,height=9cm,angle=90,clip]{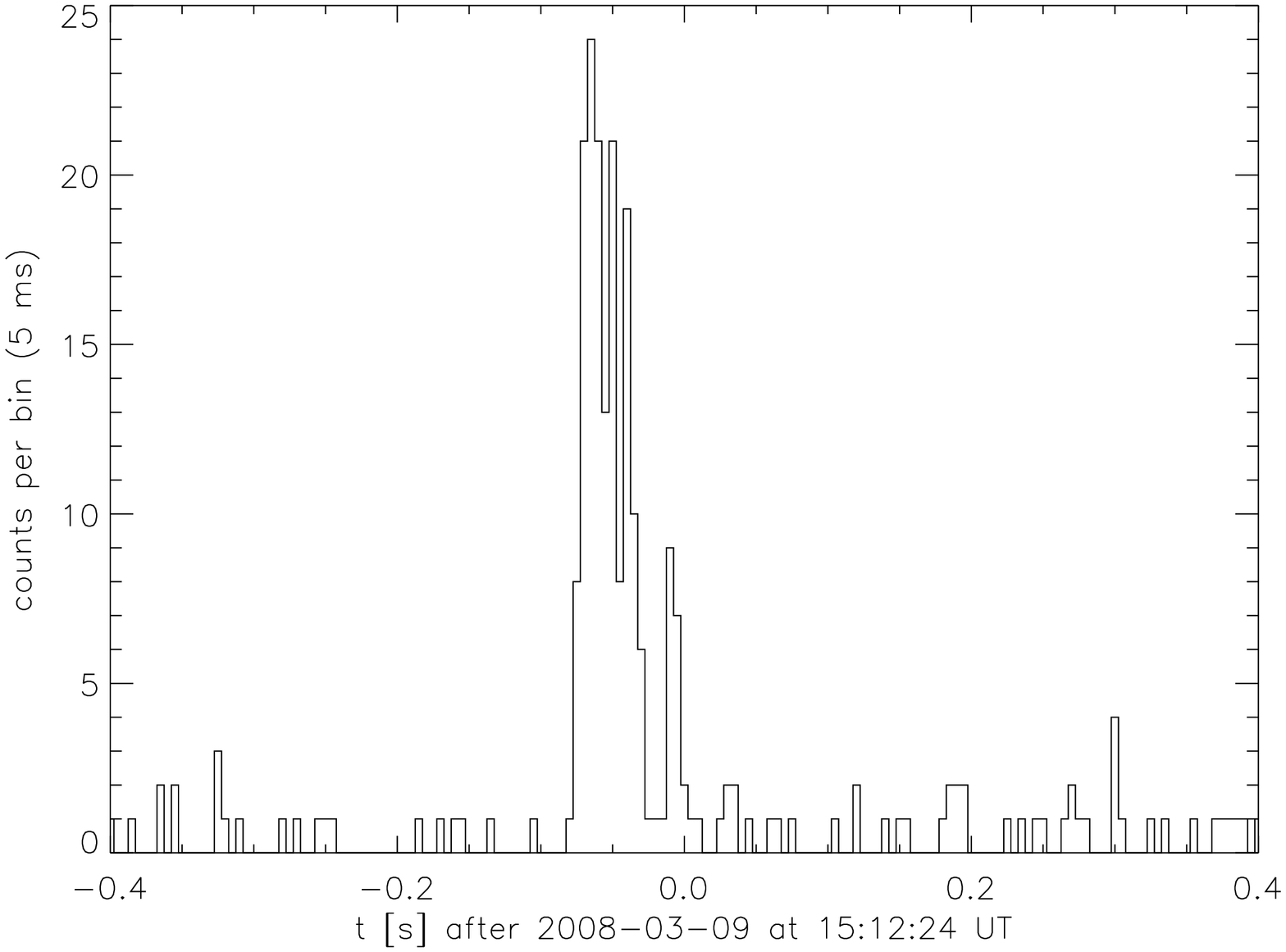}
\includegraphics[width=4cm,height=9cm,angle=90,clip]{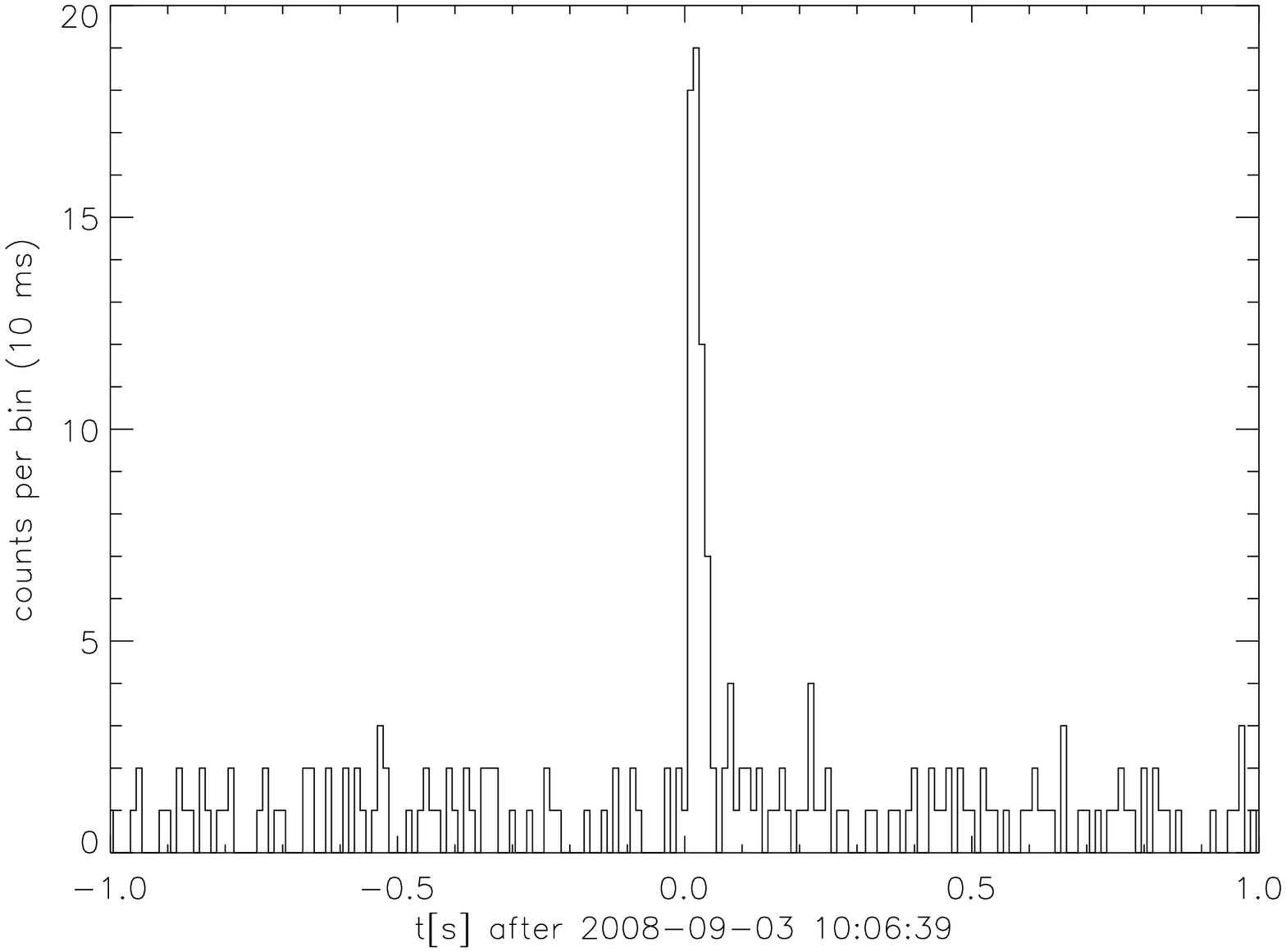}
\caption{Short bursts from SGR 1806-20 (top panel) and SGR
0501+4516 detected and localized by SuperAGILE. The time
resolution is limited by the number of detected events, but allows
to detect $<$5ms structures in the burst from SGR 1806-20. }
\label{sgr_lc}
\end{figure}

\subsection{Gamma Ray Bursts} \label{grb}

Gamma ray bursts (GRBs) are among the primary scientific
objectives for AGILE, being the first gamma-ray mission in
operation after EGRET and in the "afterglow era" of the GRB
science. Based on what known from the EGRET experience in this
field, the GRB emission in the energy range above 50 MeV is a
relatively uncommon feature in ordinary GRBs (the few percent
brightest bursts), and when present, the expected number of
photons is such that an accurate localization in the gamma-ray
range only is difficult with an experiment of $\sim$500 cm$^{2}$
effective area like the AGILE/GRID, although in case of very
bright events the smaller deadtime of GRID compared to EGRET would
favor the collection of a higher number of photons. In order to
discover the multi-wavelength GRB afterglow (e.g., soft X-rays,
optical, infrared and radio), and the source distance thereof, a
prompt arcmin-level localization of the GRID-detected GRBs is
highly desirable. SuperAGILE is in the best position in this
respect, having its field of view covering a large fraction of the
AGILE/GRID one at any time.

Indeed, SA carried out this duty quite diligently. About 1
GRB/month was discovered and localized by SA, with typical error
radii of 3 arcminutes (except for those 1-2 events localized at
the very beginning of the mission). Most of them were followed up
with the Swift X-Ray Telescope, that discovered their X-ray
afterglow, thus also indirectly confirming the accuracy of the SA
GRB positions. The first GRB localized by SA was already on 24
July 2007, during the AGILE Science Verification Phase. An
extensive report on this event may be found in Del Monte et al.
(2008a), while a description of the SA GRB triggering system may
be found in Del Monte et al. (2007).

When a bright GRB illuminates the SA detectors, during its short
duration it usually overwhelms the background count rate. In Fig.
\ref{grb_lc_image} (top panel), we show the case of one GRB,
080723B, where the peak count rate of the GRB exceeds by more than
5 times the average background rate recorded by the SA experiment.
Taking into account the source-dominated condition and the fact
that on such short time scale both the temperature and the
satellite attitude variations become negligible, images of the
event can be accumulated over the energy range from 17 to 60 keV,
deriving the projections of the source position in the SA
reference frame. They are shown in Fig. \ref{grb_lc_image} (bottom
panel) for the same event. During the GRB emission, the burst
intensity typically outshines any other source in the field. In
the few cases where a bright field source has an intensity
comparable to that of the GRB on the same timescale, that source
is usually easily identified, being among the few brightest X-ray
sources in the Galaxy (e.g., Sco X-1, or Cyg X-1, ...). From such
images an unambiguous sky position for the GRB is derived and
promptly distributed to the world wide community through the GCN
coordinate distribution network
($http://gcn.gsfc.nasa.gov/gcn3\_archive.html$).

\begin{figure}
\includegraphics[width=5cm,height=8cm,angle=90,clip]{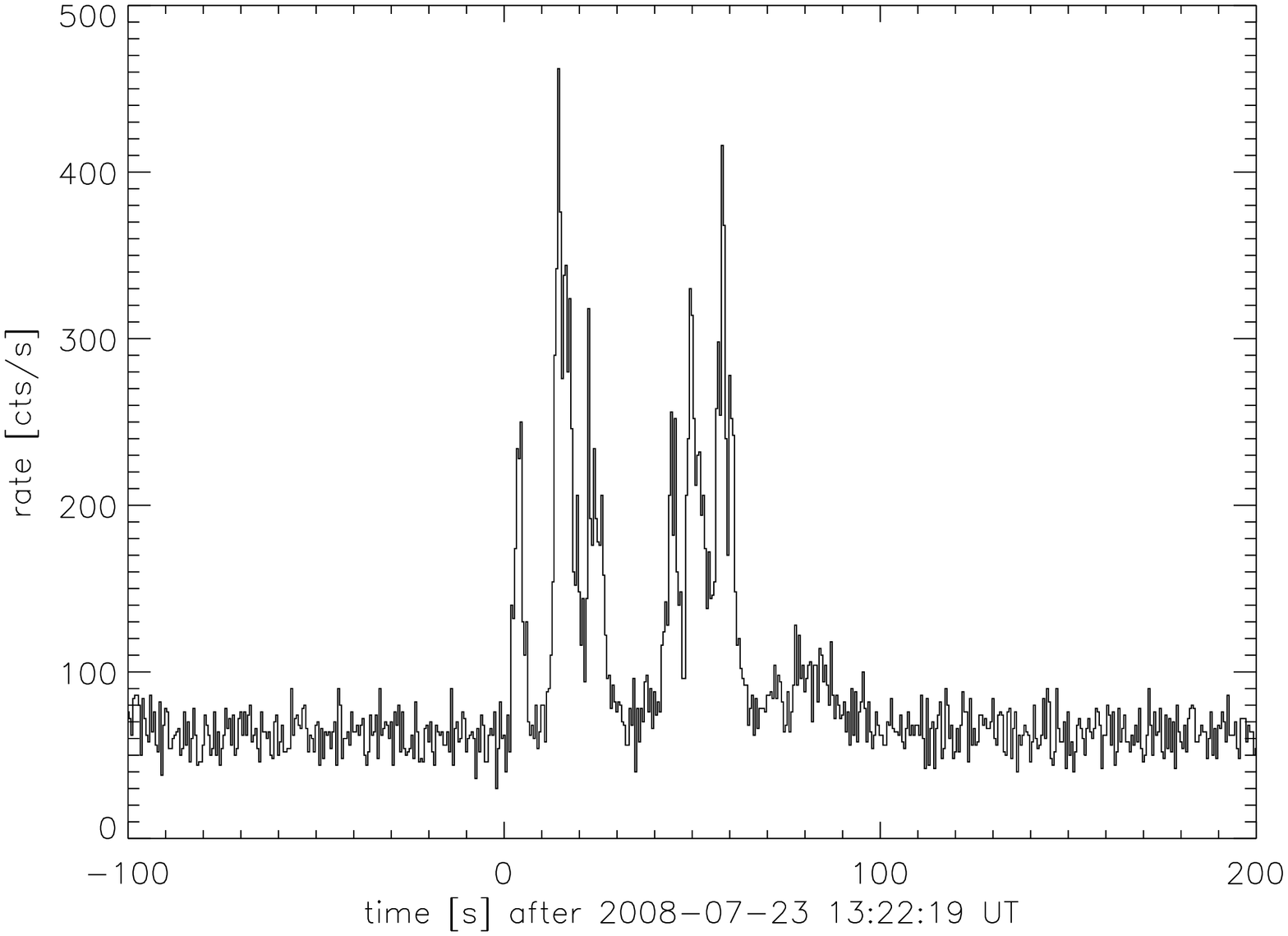}
\includegraphics[width=6cm,angle=90,clip]{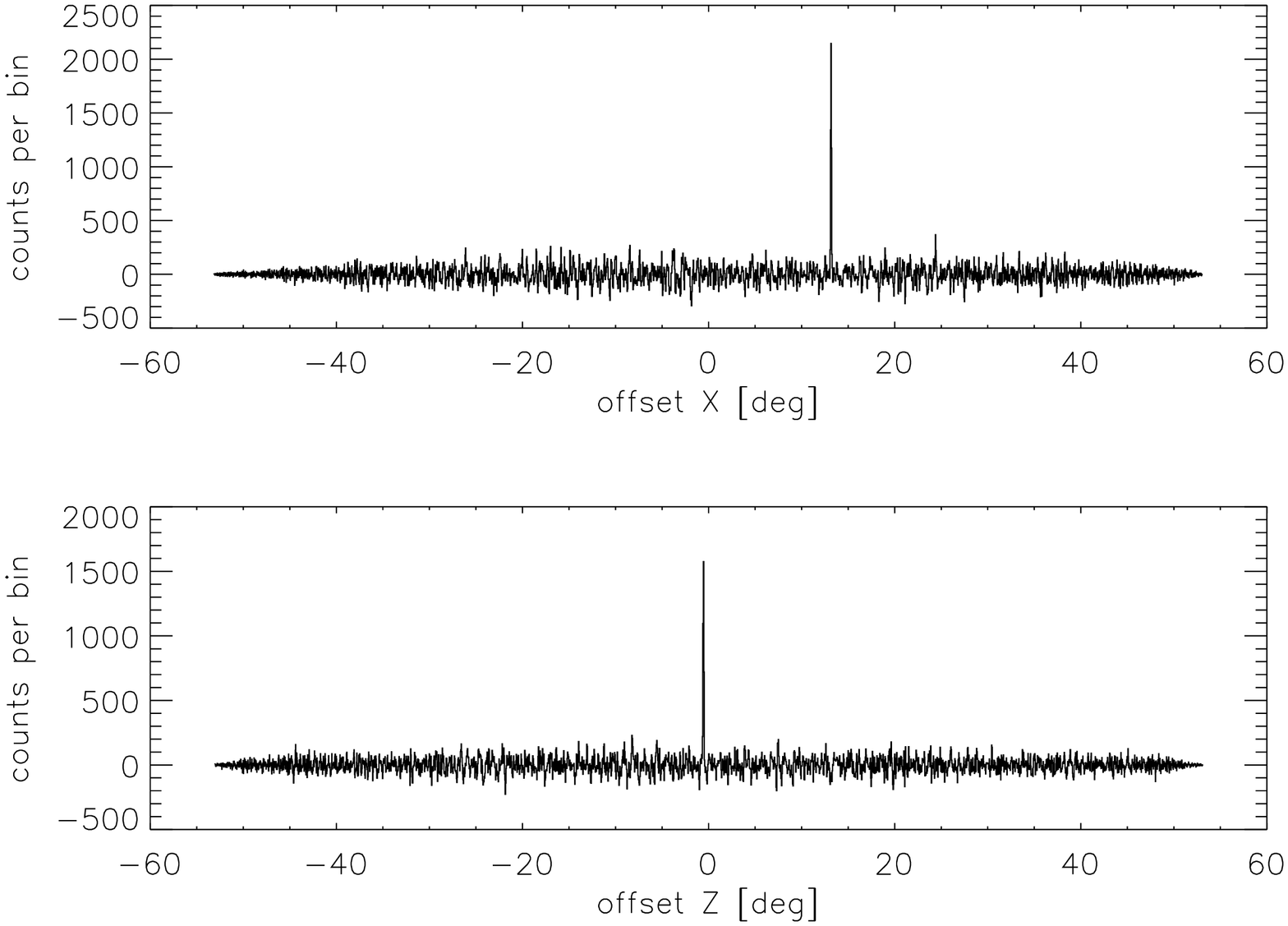}
\caption{Light curve (top panel) and images (bottom panel) of GRB
080723B, as observed by SuperAGILE. The bin size in the light
curve is 0.5 s. } \label{grb_lc_image}
\end{figure}

The case of GRB 080723B is representative of the most frequent
condition in the SA GRB detection statistics. However, in some
cases the event occurs in the part of the SA field of view with
only one-dimensional imaging encoding. In this case the SA imaging
response can only be a narrow ($\pm$3 arcmin) strip in the sky,
whose length depends on the geometry of the SA field of view and
Earth occultation constraints. Typically, it is of the order of
$\sim$10$^{\circ}$ or more. This can be reduced to the arcmin size
by including the analysis of the delay in the arrival time of the
event at the SA location and at some other long-distance
spacecraft (i.e., few light-seconds), within the context of the
Interplanetary Network (e.g., Hurley 2008). This was the case, for
example, of GRB 080514B, that was localized by SA at
X=-37.61$^{\circ}$, then in the 1D coding region, and the error
box reduced to a 100 arcmin$^{2}$ sky region thanks to a rapid
triangulation with the data of the GRB detector onboard
\textit{Mars Odyssey}. This event was especially noticeable
because it was also the first GRB detected above 25 MeV by the
AGILE GRID. It was indeed the first GRB detected at high energies
after EGRET, and the first ever GRB with emission above 25 MeV
associated with a multi-wavelength afterglow and with a
photometric red-shift \textit{z}=1.8 (Giuliani et al. 2008a, Rossi
et al. 2008).

\subsection{Extragalactic sources}
\label{agn}

The band pass and sensitivity of SA are certainly not optimal for
the study of extragalactic sources. \textbf{Only a} few bright
AGNs were detected over the reported period though (see Table
\ref{table:list}), with the noticeable cases of the radiogalaxy
Cen A, for which several detections were obtained also on the
orbital timescale, and the BL Lac source Mkn 421. The latter was
serendipitously observed during a one-week AGILE target of
opportunity observation toward the blazar W Comae. Mkn 421 was
indeed detected in a hard X-ray flaring state, with a flux peaking
at approximately 55 mCrab (Costa et al. 2008). The SA observation
of the hard X-ray flaring motivated a target of opportunity
observation with Swift/XRT that measured the source flux at its
brightest state ever at soft X-rays. An integration over the whole
week of data resulted also in a detection above 100 MeV by the
AGILE/GRID. The SA and GRID data were then analyzed in a
multi-frequency context, from optical to TeV energies (Donnarumma
et al. 2009a).

\subsection{Monitoring of AGILE GRID gamma-ray sources}
\label{grid}

The classes of persistent sources that are detectable in
gamma-rays with the sensitivity of the AGILE GRID (e.g., blazars,
gamma-ray pulsars, supernova remnants, ...) are usually distinct
from those typically detectable in hard X-rays with the
sensitivity of SA (e.g., low and high mass X-ray binaries),
although there are cases where this general scenario does not
apply, e.g. in the reported case of Mkn 421, or the Crab pulsar
(Pellizzoni et al. 2009), or the AGILE observation of the flat
spectrum radio quasar 3C 373 (Pacciani et al. 2009). However, the
SA data are systematically searched for hard X-ray excesses in
time and space coincidence with those detected by the AGILE/GRID
(and vice-versa). This resulted mostly in upper limits so far. In
some cases the SA upper limits provide significant constraints to
the interpretation of the gamma-ray data. As an example, the rapid
and bright gamma-ray transients detected by the AGILE GRID in the
Galactic plane (e.g., Longo et al. 2008; Pittori et al. 2008; Chen
et al. 2008; Giuliani et al. 2008b) are unknown in origin. They
were typically observed on the timescale of one day, and no
contemporaneous observations are available in other energy ranges.
SuperAGILE observed those sources simultaneously with the
gamma-ray detections but did not find any coincident X-ray
emission. The value of the upper limits depends on the source
position in the SA field of view and the integration time, and
ranges between 10 and 45 mCrab. The gamma-ray sources, instead,
displayed emission as large as 0.5 Crab above 100 MeV.

\section{Public distribution of SuperAGILE light curves }
\label{sa_web}

The orbital light curves of the sources detected by SA,
automatically extracted by the SASOA pipeline, provide new inputs
to the SA source database after processing of the latest available
data following every passage of the AGILE satellite over the
Malindi ground station. The  orbit-averaged source fluxes in the
standard 20-60 keV energy range are then made automatically and
publicly available to the general community, with no restrictions.
The source flux data can be accessed from the SuperAGILE web page
at AGILE Data Center
($http://agile.asdc.asi.it/sagilecat\_sources.html$) for display
and download purposes. Since they are the products of the
automatic orbital source extraction, as we discussed in Sect.
\ref{software} the detection quality is not optimized, but it is
usually good enough for the monitoring of the intensity state of
bright sources. The ASDC web page is automatically updated twice a
day, and the input of the new data into the public distribution is
then not delayed by more than 12 hours.

Orbital flux measurements are provided for only a sub-set of the
sources listed in Table \ref{table:list}, those for which
significant detections could be obtained by the orbital analysis
(identified with "Y" in the "Orbital Data" column of the table).
The brightest sources have a frequent coverage, while weaker
sources (or sources that have been only marginally observed) have
only a sparse coverage. Any source can be monitored only when it
is inside the SuperAGILE field of view. Thus, a source light curve
available from the web is usually composed of clustered sets of
measurements obtained by subsequent orbits in the observing
periods, interleaved with long gaps corresponding to periods when
the AGILE satellite was pointing at other regions. As an example,
in the top panel of Fig. \ref{sco_lc} we show the complete orbital
data set for the bright source Sco X-1 (as available from the
public web site as of April 2009), while the bottom panel shows
the same data zoomed over a one-month period (September 2008).
From the same web site the user can access the AGILE pointing plan
to know when the source of interest was or will be in the
SuperAGILE field of view.

The web page reports only the positive detections (i.e., with
significance greater than 4$\sigma$ either on the X or Z
detectors). This means that if the source flux goes below the
detection limit for one orbit, this will correspond to a missing
point in the light curve. This may also sometimes happen due to a
very small net exposure time during a given orbit, e.g. when the
source visibility (with respect to the Earth occultation) overlaps
with the satellite passage through the South Atlantic Anomaly.
These conditions can usually be recovered by a human-assisted
analysis, optimizing the source exposure. It is worth stressing
again here that being based on positive detections only, the SA
light curves currently publicly available offer an intrinsically
upward-biased view of the flux history of the sources, except for
very bright sources (i.e., above $\sim$100-200 mCrab).

\begin{figure}
\centering
\includegraphics[width=9cm,clip]{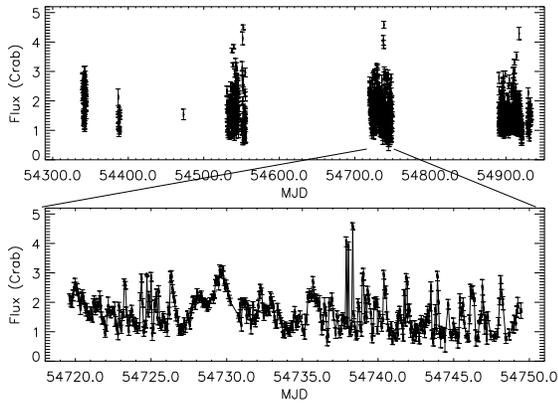}
\caption{Long-term orbital light curve of the very bright X-ray
binary Sco X-1 as it is available from the SA web page at the ASDC
web site. Top panel shows the complete data set, from July 2007 to
April 2009. Bottom panel is a zoom of a one-month time interval,
showing the short term ($\sim$hours) variability of the source. }
\label{sco_lc}
\end{figure}

Non-positive detections and upper limits can be computed as well,
typically ranging between 10 and few tens of mCrab, depending on
the integration time and the source position in the field of view.
They are currently not yet given in the web page due to computing
time constraints, but can be provided on specific sources by the
SA hardware team on request.

The automated orbital data processing applies filters to the
source detections, mostly based on the exposed area to the source
and the comparison of the source flux with a pre-set range, in
order to guarantee the reliability of the realtime, automated
source flux extraction. When a source detection is filtered out it
is automatically sent as an input to the manual processing, that
verifies the detection quality and reliability. In case of
validation the data are injected back to the public archive as
soon as the human-validated analysis is available. This implies
that, in principle, the SA source flux data are always subject to
improvements and refinements without notice, although this
condition is rather unfrequent. The temporary filter is also used
as an alert to the SuperAGILE team about special conditions of the
detected sources, in some cases leading to rapid communication
through, e.g., Astronomer's Telegrams ($
http://www.astronomerstelegram.org $).

It is useful to remind here that since the AGILE boresight drifts
by $\sim 1^{\circ}$ per day, the same source will be seen by SA at
different positions of the field of view (that is, potentially
with a largely different effective area) even during the same
pointing, having typical durations of a few weeks. A given source
may thus pass in 2-3 weeks of the same pointing from being
detectable to undetectable, or viceversa, not only due to its flux
but also to its changing position in the experiment field of view.

The SA data accessible through the web page are currently only the
results of the automatic orbital processing, described in Sect.
\ref{software}. Data products deriving from analysis on different
timescales, either shorter for bursts or outbursts, or longer for
the weaker sources, are currently not provided. Similarly, refined
source analysis are currently not provided. Work is in progress to
provide in the near future additional and more refined products,
starting with the daily integrations.

\section{Summary and conclusions}

The SuperAGILE experiment is successfully operating onboard the
AGILE mission since 2007, April 23$^{rd}$. In this paper we
provided a description of the SA data, and reported an overview of
the main scientific results achieved in the first $\sim$20 months
of scientific observations. The main goals of SuperAGILE are the
simultaneous hard X-ray observation of the central region of the
field of view of the AGILE/GRID experiment, the prompt discovery
and localization of gamma-ray bursts and Galactic transients, and
the long-term monitoring of the bright Galactic sources.

The aim of the simultaneous hard X and gamma-ray observation of
the same field is to discover correlated variability of sources in
these two energy ranges, and use the finer angular resolution of
SuperAGILE to localize gamma-ray sources detected in the GRID.
This is the first time that an X-ray and a gamma-ray imager
systematically observe the same field simultaneously. Regrettably,
this did not bring yet to discover any previously unknown
correlated behavior of sources. The sources that SuperAGILE and
GRID detected simultaneously were already known to emit in both
energy ranges. This is the case for the Crab pulsar, or the AGNs
3C 273 and Mkn 421. However, the guaranteed simultaneous GRID and
SuperAGILE observations represent the basic seed of multifrequency
campaigns, often complemented with radio to TeV observations. This
led, for example, to detect a flaring state of the BL Lac source
Mkn 421 and interpret the time variability of the simultaneous
optical-to-TeV spectral energy distribution in terms of a rapid
acceleration episode of the leptons in the jet (Donnarumma et al.
2009).

The primary goal is of course to search for positive detections in
both instruments, but there are cases where also upper limits in
one or the other are important to interpret the origin of the
detected emission. This was the case of the unidentified transient
sources discovered by the AGILE/GRID on the galactic plane, in the
Cygnus and Musca regions (e.g., Longo et al. 2008; Pittori et al.
2008; Chen et al. 2008; Giuliani et al. 2008b). Following Romero
\& Vila (2009), the SuperAGILE upper limits on the simultaneous
hard X-ray emission of these transient gamma-ray sources imply a
very large ratio between the gamma-ray and X-ray luminosity,
posing severe constraints on their interpretation in terms of
emission from jets of galactic microquasars, strongly favoring a
dominant hadronic component in the jet.

Interesting results were obtained also in the field of gamma-ray
bursts. Despite of the prompt SuperAGILE arcmin-localization of
several GRBs, enabling searches in time and spatial coincidence,
in the reported period only two events were significantly detected
by the AGILE/GRID (Giuliani et al. 2008a, Moretti et al. 2009),
including GRB080514B, the first GRB ever for which the gamma-ray
emission can be correlated to an afterglow counterpart with a
measured distance, and GRB 090401B. The detection of these events
by all the AGILE instruments - SA (18-60 keV), GRID ($>$25 MeV),
MCAL (350-700 keV) and ACS ($>$8 keV) - also allowed to measure a
few seconds delay of the emission above 25 MeV with respect to
that at lower energies. The observations of the \textit{Compton
Gamma Ray Observatory}, particularly the EGRET and BATSE
experiments, provided the result that the GRBs with a gamma-ray
counterpart detectable above the EGRET sensitivity (that is
similar to AGILE/GRID's) belong to the brightest 5\% of the BATSE
GRB distribution, but the statistics was only that of 5 events.
Although the SuperAGILE GRBs are selected towards the bright end
of the BATSE distribution, because of its smaller area, the 2 GRID
detections over 21 GRBs localized by SuperAGILE by the end of
April 2009 are still consistent with the approximate EGRET
statistics.

The pointing strategy of the AGILE mission, typically consisting
of long exposures, offers as a by-product a long term monitoring
of the same field by SuperAGILE and as a drawback a very
inhomogeneous sky coverage (see Fig. \ref{fig_exposure}), largely
privileging the Cygnus and Vela sky regions in particular.
Although the analysis of the SuperAGILE data collected so far is
not complete yet (see $\S$\ref{software} and
$\S$\ref{detected_sources}), about 60 sources were detected over
the time period reported in this paper. As it can be seen from
Table \ref{table:list}, the large majority of them are X-ray
binaries, mostly Low Mass X-ray Binaries (LMXB).

Despite the inhomogeneity and incompleteness of the SuperAGILE
exposure and data analysis, it may be useful to put the list of
sources detected by SA in the context of the current scenario in
the hard X-rays. Currently, INTEGRAL/ISGRI and Swift/BAT offer the
most complete sky surveys in the hard X-rays. The list of sources
detected by SuperAGILE given in Table \ref{table:list} was
obtained mainly by daily integrations. The sensitivity of
SuperAGILE on this timescale is in the range of $\sim$15-20 mCrab.
Applying a flux cut at 15 mCrab in the ISGRI 3.5-year catalogue
(20-40 keV, Bird et al. 2007) and to the BAT 22-month catalog
(14-150 keV, Tueller et al. 2009) we found about 50 and 40
sources, respectively. Of these, in ISGRI 38\% are HMXB and 57\%
are LMXB, while in BAT they are 31\% and 59\%, respectively. From
the list in Table \ref{table:list}, SA detected 57 persistent
sources, of which 33\% are HMXB and 58\% are LMXB. As expected, a
large number of sources are actually in common, being the bright
tail of the LogN-LogS distribution of persistent hard X-ray
sources. Although the agreement is remarkably good, the above
comparison should not be taken rigorously, due to a large number
of important caveats about pointing strategy, type of the
analysis, exposure, energy range and intrinsic source variability.
However, it is a nice general confirmation that the source
detection record by SuperAGILE is consistent with the sensitivity
limit of the current analysis.
\\
Pushing the comparison (as well as the caveats) beyond, we made
the same type of selection to the source catalog of the
BeppoSAX/Wide Field Cameras (WFC, Verrecchia et al. 2007). This
experiment operated in the 2-26 keV energy range. Selecting the
sources with reported average flux above 15 mCrab in 2-10 keV, we
found $\sim$80 sources, of which 20\% are HMXB and 73\% are LMXB.
A comparison between the type and number of X-ray binary sources
brighter than 15 mCrab below $\sim$20 keV (in the WFC) and above
(in SA, BAT and ISGRI) shows that approximately the same number of
objects is found among the HMXB, while the LMXB decrease to one
half above 20 keV. This suggests that the LMXB, as a class, tend
to have energy spectra softer than the Crab (photon index
$\sim$2.1), while the HMXB tend to have harder spectra, on
average.

Many of the SA sources were detected on the orbital timescale,
meaning that public light curves are available for them at the
ASDC web page (see $\S$\ref{sa_web}), updated daily with new data
(when available). In this paper we provided detailed explanations
on how these light curves are extracted and under what
assumptions, thus offering an understanding of the limits of their
scientific use. Other sources have been detected either in longer
integrations or in short outbursts. For these sources no publicly
accessible science products are currently available yet.

The main characteristics of the SA observation is the continuity
and long duration. Some galactic sources have been observed for
probably the longest continuous stretches of time ever, allowing
for monitoring the long term variability of their hard X-ray flux.
The duration and continuity of the monitoring distinguishes the SA
data from those of more sensitive experiments (e.g., Swift/BAT)
that offer short and sparse observations of individual sources,
sometimes missing short-term variability or events. In addition,
the SA field is always simultaneously observed by the AGILE/GRID,
thus offering the simultaneous measurement of the flux above
$\sim$50-100 MeV.

The SuperAGILE experiment is operating nominally since its
switch-on in orbit in 2007, showing no signs of degradation or
losses with time.
The in-flight operation of the AGILE mission
is expected to continue until at least mid-2011, subject to
approval by the Italian Space Agency (ASI).

\begin{acknowledgements}
AGILE is a mission of the Italian Space Agency (ASI), with
co-participation of INAF (Istituto Nazionale di Astrofisica) and
INFN (Istituto Nazionale di Fisica Nucleare).
\end{acknowledgements}

\end{document}